# Multiscale Analysis of Plasma-Modified Silk Fibroin and Chitosan Films

Jordan Nashed*[a,b], Tomasz Bartkowiak[c], Alexandru Horia Marin[d,a,f], Tine Curk[b,g], and Viviana Marcela Posada-Perez*[a,e]

[a]Ken and Mary Alice Lindquist Department of Nuclear Engineering, Pennsylvania State University, PA USA

[b]Department of Materials Science and Engineering, Johns Hopkins University, MD USA

[c]Institute of Mechanical Technology, Poznań University of Technology, plac Marii Skłodowskiej-Curie 5, 60-965 Poznań, Poland

[d]Materials Science and Engineering Department, Pennsylvania State University, University Park, PA, 16802, USA

[e]Department of Electronics and Computer Science, Pontificia Universidad Javeriana de Cali, Colombia

[f]Surface Analysis Laboratory, Institute for Nuclear Research Pitesti, Mioveni, 15400, Romania

[g]Department of Physics and Astronomy, Johns Hopkins University, MD 21218

*Corresponding authors

## Abstract:

This study establishes a multiscale surface characterization framework to identify biologically relevant topographic length scales on plasma-modified polymer surfaces. Biological interactions with material surfaces span a wide range of length scales, yet most surface characterization approaches rely on single-scale metrics, limiting their ability to explain scale-dependent biological responses. Multiscale surface descriptors of plasma-modified silk fibroin and chitosan surfaces were correlated with bacterial and immune cell responses. Surface chemistry and topography were characterized using X-ray Photoelectron Spectroscopy (XPS) and Atomic Force Microscopy (AFM), followed by sliding bandpass filtration and curvature tensor-based analyses to quantify scale-dependent topographic features. Macrophage response and biofilm growth were assessed by fluorescence microscopy. Correlation analyses revealed strong scale dependence: individual bacteria and small colonies correlated primarily with fine-scale topographic features, whereas macrophage morphology correlated more strongly with larger-scale surface features. In contrast, surface chemical descriptors showed weak correlation with biofilm formation, despite clear differences in bacterial support between chitosan and silk fibroin, suggesting that material identity is not fully captured by standard chemical metrics. Together, these results demonstrate that multiscale topographic analysis enables identification of surface feature sizes most relevant to specific biological interactions, supporting its use as a surface engineering tool for functional biomaterials.

## 1. Introduction:

Control of biofilm formation and inflammatory response on material surfaces is critical for maintaining functional performance of biomaterials and biomedical devices, motivating surface design strategies that suppress bacterial adhesion while maintaining compatibility with host cells [1,2]. Silk fibroin (SF) [3–8] and chitosan (Ch) [9–11] are widely studied polymeric substrates whose interfacial properties can be tailored through surface modification to regulate biological interactions [7,12,13]. In particular, physically inspired surface patterning strategies derived from naturally bactericidal surfaces have attracted interest as chemical-free approaches to bacterial control [14–16]. Although the underlying bactericidal mechanisms remain under debate [17], high-aspect ratio nanostructures are widely reported to mechanically disrupt bacterial membranes without inducing cytotoxic responses in mammalian cells [18,19].

Our group has explored the use of a surface modification technique known as “Directed Plasma Nanosynthesis” (DPNS) as a means of creating high aspect-ratio nanostructures on biomaterial surfaces [7,13,20–22]. This technique extracts ions from a plasma source and accelerates them towards a target material at low energies (~100-500 eV) and a user-selected incidence angle, enabling the creation of oriented nanostructures on material surfaces [7,20]. Importantly, the structures created by DPNS are more complex [7,13] than the highly-controlled patterns typically used in the study of bactericidal topographies [23–26], presenting a metrological challenge for analyzing them quantitatively.

Biological interactions with material surfaces occur across multiple length scales, from nanometer-scale protein adsorption that conditions subsequent cell adhesion to micron-scale bacterial attachment and cellular organization [27,28]. Proteins, which are typically on the nanometer scale, are primarily influenced by nanoscale topography [29], whereas bacteria (~1 μm) and individual eukaryotic cells (~10 μm) respond to both nano- and microscale surface features [13,24,26]. In contrast, collective cellular organization and colony-scale behavior are influenced by larger-scale surface structure (>100 μm) [28]. This inherent multiscale coupling between surface features and biological response motivates the use of scale-discriminant (multiscale) surface analysis methods.

Multiscale analysis in surface metrology can be defined as "the process of studying topographies at multiple scales of observation and comparing, merging, or associating the findings acquired from observations or calculations at different scales" [30]. This can be achieved in a number of ways, from spatial filtering to taking images at different length scales or spatial resolutions [30]. It may also involve finding the surface characterization parameters whose geometric properties are best discerned at the specific scale or scales of observations or identifying the scale or scales in which interactions between formation process and resulted surface topography or between surface topography and functional response occur [31]. Despite their relevance, multiscale techniques remain underutilized and poorly standardized in biomaterial surface studies; instead of considering the full range of surface length scales, much of the field continues to debate whether nano- or micro-scale topography dominates biological responses [32–37]. Additionally, while qualitative descriptors of surface topographies can be useful, a more systematized, quantitative approach to characterizing surface topographies would enable more consistent comparison between studies of surfaces of arbitrary complexity [29,30].

The clinical relevance of silk fibroin (SF) and chitosan (Ch), together with their inherent limitations and the emergence of directed plasma nanosynthesis (DPNS) as a physical surface modification strategy, motivated our previous investigations of DPNS-treated SF [7] and Ch [13]. Building on this work, the present study examines how plasma-induced surface modifications influence biological interactions by correlating surface descriptors with bacterial and macrophage responses across relevant topographic length scales. Two complementary multiscale approaches were applied to Atomic Force Microscopy (AFM) topography data to resolve scale-dependent surface features. The first, developed by Bartkowiak et al., employs multiscale curvature tensors to estimate local surface geometry across multiple observation scales [38]. The second, a bandpass filtration approach, isolates topographic features within prescribed length-scale windows and was originally introduced for multiscale surface analysis by Berglund and Brown [39,40]. Conventional single-scale (as-measured) topographic descriptors were also included to assess whether biologically relevant surface features are overlooked when scale is not explicitly considered. Surface chemistry, characterized by X-ray photoelectron spectroscopy (XPS), and topographic parameters were treated as explanatory variables, while bacterial adhesion metrics, biofilm size distributions, and macrophage morphological descriptors were used as biological response variables. This framework enables identification of the surface feature sizes most strongly associated with distinct biological responses and allows direct comparison between multiscale and traditional surface descriptors for plasma-modified SF and Ch surfaces.

## 2. Methodology

### *2.1 Film Synthesis*

SF films were obtained following a standard LiBr-based extraction workflow [7,41]. Briefly, *Bombyx mori* cocoons were degummed in 0.02 M sodium carbonate to remove sericin, rinsed thoroughly, and dried. The resulting fibroin was dissolved in 9.3 M lithium bromide at 60°C and subsequently dialyzed (3.5 kDa Molecular weight cut-off) against ultrapure water for 48 h with multiple water exchanges. The clarified SF solution (≈7.8 wt/vol%) was then concentrated by osmotic dialysis against a 10% polyethylene glycol solution (PEG 10,000), yielding a viscous SF solution of ~15.9 wt/vol%. Films were formed by casting this concentrated solution onto glass substrates and allowing it to dry under ambient conditions. For Ch films, a 1% (w/v) Ch solution (≥75% deacetylation, low molecular weight) was prepared in 1% (v/v) acetic acid using sonication and magnetic stirring. The solution was cast onto 10 × 10 mm glass substrates (200 µL per slide) and dried overnight at room temperature in a flow hood. The dried membranes were neutralized in 5% (w/v) sodium hydroxide for 1 h, rinsed to neutral pH, and dried again at 37 °C to obtain the final Ch films [13].

### *2.2 DPNS*

SF and Ch films were placed in a vacuum chamber evacuated to a base pressure of $5 \times 10^{-6}$ Torr and subsequently filled with argon gas to reach a working pressure of $3 \times 10^{-4}$ Torr. Surface modification was performed using argon ions ($Ar^+$) generated from plasma at an energy of 1 keV, with a beam flux of $1.5 \times 10^{15}$ $cm^{-2} \cdot s^{-1}$. Following this, a second irradiation step was applied using $Ar^+$ ions at 500 eV with a fluence of $1 \times 10^{18}$ ions/$cm^2$. Treatments were conducted at incidence angles of 45° and 60°, and the resulting samples were designated as SF 45, SF 60, Ch 45, Ch 60, and their respective pristine controls (SF Control and Ch Control).

Under the low-pressure conditions in this work (~$3 \times 10^{-4}$ Torr), ion trajectories are governed by sheath-accelerated electric fields, resulting in predominantly anisotropic ion bombardment of the substrate. At this pressure, the ion mean free path (~20 cm) is significantly larger than the few-centimeter source–substrate working distance [42], enabling predominantly ballistic transport with negligible probability of gas-phase collisions. This advocates for the reproducibility of surface modifications made by DPNS. The defined ion energy, flux, fluence, and incidence geometry ensure controlled momentum and energy transfer to the near-surface region, driving nanoscale restructuring through ion–surface interactions. A similar experimental configuration for studying ion–surface interactions under controlled vacuum conditions has been reported for the IGNIS-2 plasma–material interaction facility [43].

### *2.3 Surface characterization*

For SEM analysis, cell-seeded samples were fixed, dehydrated through a graded ethanol series, stored overnight in a desiccator, and sputter-coated with iridium. Cell-free samples were cleaned, dried, and sputter-coated directly. All samples were imaged using a Thermo Scientific Verios G4 SEM or a Thermo Scientific Scios FIB under high-resolution secondary electron detection.

XPS investigations were performed using a Physical Electronics VersaProbe III instrument equipped with a monochromatic $AlK_\alpha$ X-ray source (hv=1486.6 eV, 200 μm spot size) and a concentric hemispherical analyzer. The emission angle was 45°. Charge neutralization was achieved with a dual-beam charge neutralization system (low-energy electrons and low-energy $Ar^+$ ions). Before the experiments, the binding energy scale of the spectrometer was calibrated by setting the Fermi edge of sputtered gold to 0 eV and the $Au4f_{7/2}$ line to 84.0 eV. The acquired high-resolution spectra were charge referenced by fixing the C1s peak of aliphatic carbon at 285.0 eV [44]. Measurements were made at a takeoff angle of 45 with respect to the sample surface plane.

Five AFM images per material-treatment group were obtained from distinct SF and Ch films, resulting in thirty independent images. Images were obtained using Bruker's Icon Dimension II or Bruker's Icon AFM-IR, both in PeakForce tapping mode. Images were all acquired at 5×5 μm and 512×512-pixel resolutions to enable consistent comparison and multiscale analysis between all images.

### *2.4 Bacterial Growth*

Escherichia coli (*E. coli*) was cultured in tryptic soy broth at 37°C under mild agitation (50 rpm) for 24 h. The culture was subsequently diluted to a final concentration of ~26 × $10^6$ CFU/mL. Each sample was exposed to 2 mL of the bacterial suspension and incubated at 37 °C for 1 h under static conditions. Following incubation, nonadherent bacteria were removed by rinsing samples with phosphate-buffered saline. The remaining surface-bound bacteria were stained using SYTO 9. Fluorescence imaging was performed using a confocal microscope equipped with FITC (excitation/emission: 480/500 nm) filter. For each surface condition, two independent samples were analyzed, with three spatially distinct regions imaged per sample. In another set of samples bacteria were fixed and dehydrated with consecutive ethanol dilutions in water. The samples with the bacteria cells were coated with iridium and SEM images were obtained using a Verios G4.

Bacterial proliferation was quantified in ImageJ via statistics gathered from ImageJ's "particle analyzer". These statistics are the fractional area of the image covered by bacteria (denoted as area coverage) and the area density of small, medium, and large bacterial colonies, which were defined as falling between 1-20 $\mu m^2$, 20-200 $\mu m^2$, and > 200 $\mu m^2$, respectively. This approach was taken due to the extreme skewness of the colony size distribution (**Supplementary Figure S2**) as well as the biological motivation to observe the differential effect of DPNS on individual bacteria (i.e., small colonies) and large biofilms. These bacterial statistics were normalized by the area of the analyzed region. Additionally, medium and large colony circularity were assessed via ImageJ's particle analysis tool to derive semi-quantitative measures of colony EPS production and maturity [45].

These data were obtained by first selecting a square region of interest (ROI) from a fluorescence image, subtracting the background using 50.0 pixel rolling ball radius with sliding paraboloid without smoothing before applying a threshold, whose value was determined visually such that none of the observed bacteria were removed. After thresholding, ImageJ's particle analyzer created a ROI which could be pasted onto the original grayscale 8-bit image for subsequent measurement. A list of particle sizes was then obtained along with the total area covered by the identified particles.

### *2.5 Macrophage morphology*

Samples were first exposed to UV light for 20 min for sterilization. J774 murine macrophages were then seeded onto the samples at a density of 20,000 cells per sample for immunofluorescence staining. Cells were initially deposited in droplets of 30–50 µL and allowed to attach for 1 h. After attachment, the medium was supplemented with 650 µL of Dulbecco's Modified Eagle Medium (DMEM) containing fetal bovine serum (FBS).

Macrophages were fixed with 4% paraformaldehyde, permeabilized with 0.1% Triton X-100, and blocked with 2% bovine serum albumin (BSA). Cells were incubated overnight at 4 °C with rabbit anti-CCR7/CD197 (Bioss, bs-1305R) and rat anti-CD206 (MR5D3; Invitrogen, MA5-16871), following a standard secondary-detection protocol, with buffer-only controls included. After three PBS washes, samples were incubated with FITC-conjugated goat anti-rabbit IgG (Invitrogen, A16118) for CCR7 and Alexa Fluor™ 633 goat anti-mouse IgG2a (Invitrogen, A-21136) for CD206. Nuclei were counterstained with DAPI, and samples were imaged under fluorescence microscopy. Merged fluorescence masks were used to quantify macrophage coverage and clustering. Clusters were defined as spatially contiguous macrophage-positive regions (connected components) within the merged mask, and ImageJ's particle analyzer was used to derive area distribution statistics from macrophage clusters. Macrophage circularity was determined following the same procedure as for clusters, but with the addition of the watershed algorithm prior to particle analysis to identify and analyze individual macrophages.

For semiquantitative analysis, images were processed in ImageJ (NIH). Background subtraction was applied, and fluorescence intensities for CCR7 and CD206 were measured within segmented macrophage regions. Marker intensities were corrected by their respective exposure times to account for acquisition differences. Normalized CCR7 and CD206 values were then calculated as the relative contribution of each marker to the total fluorescence signal within the same image, yielding complementary fractions that sum to unity.

Merged fluorescence masks were used to quantify macrophage coverage and clustering. Clusters were defined as spatially contiguous macrophage-positive regions (connected components) within the merged mask, and ImageJ's particle analyzer was used to derive area distribution statistics from macrophage clusters.

### *2.6 Multiscale Topographic Analysis*

#### *2.6.1 Image Processing*

Raw AFM images were initially processed in Gwyddion 2.63. First, mean plane subtraction was applied. If scan line artifacts or scars were present, Gwyddion's "align rows" tool was implemented using polynomial of degree one followed by the "correct horizontal strokes" tool. All images were adjusted to set the minimum height value to 0 nm before further processing. A MATLAB script was written to convert files from Gwyddion into the format needed for multiscale curvature analysis in Mathematica. The MATLAB and Mathematica scripts are available upon request, and pseudocode for the Mathematica script is available in [46].

### *2.6.2 Multiscale Curvature Tensor Method*

Text files of the processed images were analyzed using the Mathematica script produced by Bartkowiak et al. [38], wherein each AFM image was analyzed at eight different scales increasing linearly from ~10 nm to ~0.5 μm. These scales are distinct from those in the bandpass filtration method due to the fundamental difference in how scale is defined in each method. A schematic depiction of the curvature computation is shown in **Figure 1a**. From the multiscale tensor method, 15 parameters related to surface curvature were computed and analyzed. These fifteen parameters were grouped into two families, complexity, and signed curvature. Complexity refers to changes to topographic parameters with scale as well as irregularity/changes in curvature. Signed curvature is the average 'character' of a surface curvature; positive signed curvature indicates higher peak character while negative signed curvature indicates increased valley character. Individual variables, their definitions, and assigned families are described in the (**Supplementary Information, Appendix A**).

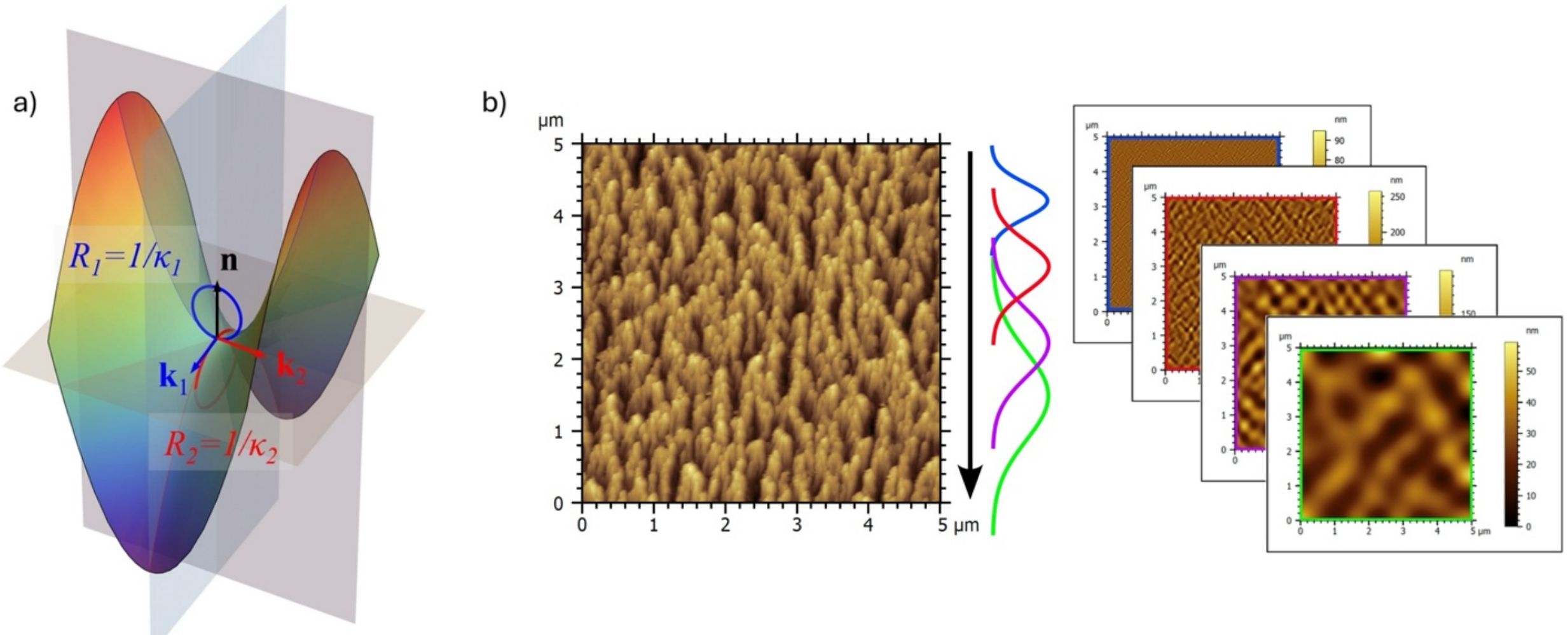

**Figure 1. Multiscale methods description.** a) Schematic representation of the principal curvatures k1 and k2, calculated at an example saddle-like location, used in the multiscale tensor

method. b) Example of the sliding bandpass filtration approach, showing the raw AFM topography and a series of filtered images at different scales.

#### *2.6.3 Sliding Bandpass Filtration Method*

Sliding bandpass filtration was applied in MountainsMap 11 [47] Free Trial software using the 'bandpass filter bank' operator with first-order ISO Gaussian filters, two bands per octave, and automatic cutoffs. This effectively resulted in fourteen images – one unfiltered, and thirteen filtered – for each of the thirty raw AFM images acquired. Sliding bandpass filtration is shown in **Figure 1b**. From each of these images, 29 ISO standard parameters [48] were collected to holistically characterize the surface topography. For ease of analysis and interpretation, these parameters were grouped into ten different families: Anisotropy, Void Volume, Hill Volume, Void Area, Hill Area, Projected Area, Roughness, Complexity, Peak Curvature, and Void Curvature (**Supplementary Information, Appendix A**). Scale here is interpreted as the central wavelength of the bandpass filter.

#### *2.6.4 Single-scale Analysis*

Single-scale analysis was performed by calculating the same ISO parameters and families used in the bandpass method on unfiltered AFM images.

#### *2.6.5 Analyzing Topographic Changes*

AFM images were examined to gain insight into the morphological changes induced by plasma treatment (**Figure 2a-f**). These qualitative observations were supplemented by plotting several multiscale topographic parameters as a function of scale (**Figure 2g-l**).

#### *2.6.6 Visualizing Group-Level Surface Changes: PCA*

Principal Component Analysis (PCA) was performed on the single-scale topography and XPS data as well as on the multiscale topography and XPS data to visualize group-level differences in material surface properties. Each of the thirty total samples' scores were plotted on PC1 and PC2 to visualize their grouping in latent space (**Supplementary Figure S4**). Since PC1 captured most of the variance in material data and was most discriminatory with respect to material and treatment for both single-scale and multiscale data, its loadings were investigated explicitly to better understand the latent structure in the surface data. To accomplish this, the top two largest loadings per family were visualized (**Supplementary Figure S5**).

### 2.6.7 *Visualizing Correlation with Biological Data: Heat Maps*

Heat maps were created to visualize the correlation of material parameters with biological response as a function of surface chemistry, topographic feature, and, in the case of multiscale data, scale of observation. Chemical data was first normalized to ensure fair comparison between material and treatment conditions. This involved taking the chemical state relative concentrations (**Tables 1 and 3**) and multiplying them by their respective element relative concentrations (**Table 2**) for each material and sample to normalize for atomic concentration. The analyzed functional groups from the C1s peak were C-C/C-H, C-N/C-O/C-$NH_2$, C-N-C=O/C=O, R-O-C=O/C-$N_3$, and pi to pi* shakeup; from the N1s peak, $NH_3^+$ was analyzed. For topographic data, parameters were sorted into several families (**Supplementary Information, Appendix A**), where a particular family represented a shared characteristic captured by its constituent parameters. This family-level grouping was employed for several reasons: 1) variables within families often had very strong collinearity, making separation of variables within a family redundant and uninformative, 2) interpretation was vastly improved by this dimensionality reduction, and 3) grouping variables from a given family helped mitigate spurious correlation. Bandpass and multiscale curvature tensor data were separated due to the differences in their respective scale measurements.

Data for each feature was mean-centered and reduced to unit variance (i.e., Z-scored) to normalize units. Each sample data was then averaged within each family (and scale bin, in the case of multiscale data) to give a sample-level score per family. Sample scores were group-wise averaged to give a group-level (i.e., SF 45, SF Control, etc.) score for a particular family and scale. Biological data was group-wise averaged for determination of Spearman’s rho in subsequent steps. Spearman’s rho was selected as the correlation metric because it strictly is concerned with monotonicity in the relationship between material and biological data (as opposed to the linearity, as in Pearson’s r). If a given family contained no variables after the statistical significance screening, its cells were greyed out in the final heatmap. To mitigate spurious correlation in the multiscale data, if correlation values for a given family differed by more than 1.0 at adjacent scales of observation, those cells were similarly greyed out.

# 3. Results and Discussion

## *3.1 Impact of DPNS on Material Surface Properties: Observations from Raw Data*

### *3.1.1 Topographic Changes*

Most nanoscale studies, particularly those correlating biological variables with surface features, rely on single-scale topographic descriptors [49,50]. Given that biological interactions occur across multiple length scales, surface modification strategies capable of generating features at different scales are required to probe scale-dependent effects. In this work, $Ar^+$ irradiation was used to induce topographic modifications while introducing only mild changes in surface chemistry.

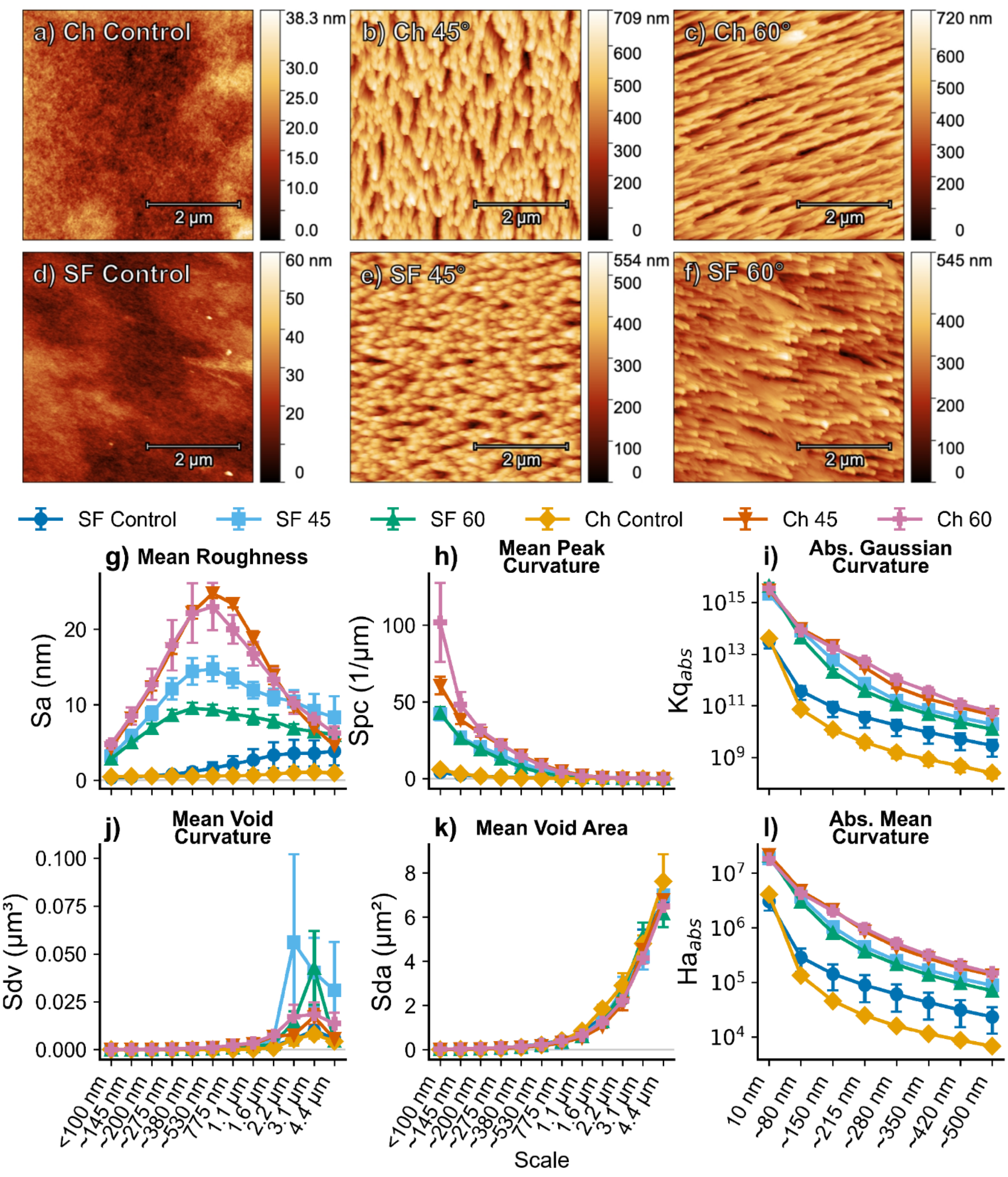

**Figure 2. Angle-induced topography changes in Ch and SF surfaces.** 5×5 µm AFM images for chitosan (Ch) and silk fibroin (SF) under control conditions and after treatments at 45° and 60° (a–f). The images show how the incidence angle drives changes in surface morphology and nano-feature organization. Panels (g–l) present the quantified surface-texture parameters: g) mean

roughness (Sa), h) mean peak curvature (Spc), i) standard deviation in absolute Gaussian curvature ($Kq_{abs}$), j) mean void volume (Sdv), k) mean void area (Sda), and l) average absolute curvature ($Ha_{abs}$) as a function of scale (note: bandpass scale axis is nonlinear due to binning algorithms). The curves compare the effect of treatment angles across material surfaces.

Irradiation created major topographic alterations in both materials, as shown in the AFM scans in **Figure 2 (a–f).** Off-incidence ion irradiation produced an oriented "grass-like" morphology, as previously observed in our reports and in the literature [7,13,51]. The anisotropy of these structures increased with an irradiation angle, again consistent with previous results [7,13]. Multiscale analysis showed that increasing the irradiation angle reorganized surface complexity rather than uniformly amplifying roughness, shifting curvature density toward smaller scales (**Figure 2h**).

Key multiscale parameters extracted from the AFM scans were plotted as a function of scale in **Figure 2 (g- l)**. Notably, the average roughness (Sa) peaks at intermediate scales, from 200 nm - 1.6 μm, giving an estimate for the scale at which the tallest structures (i.e., the oriented nanostructures) were formed. Treated Ch showed higher roughness than treated SF, irrespective of irradiation angle. Mean peak curvature (Spc) increased with decreasing measurement scale for all treated materials, and was largest for treated Ch, indicating a high degree of nano-scale asperities introduced by irradiation. Mean void volume (Sdv) was expectedly largest at the largest scales and was highest for treated SF. Mean void area (Sda) was highly consistent among all six groups and decreased with scale.

Single-scale data (**Supplementary Figure S1**) corroborated the above findings, though lacking the added depth provided by multiscale analysis. Single-scale measurements showed that Ch was rougher, but multiscale analysis revealed that this roughness occurred primarily from ~300 nm - 1 μm, near the characteristic scale of individual bacteria [25]. Treated Ch exhibited the highest average roughness, peak curvature, and void volume. Control SF, on the other hand, had the largest mean void area (Sda), likely due to the identification of fewer, larger voids in SF Control compared to Ch Control, as determined by the watershed algorithm used to compute Sda.

### *3.1.2 Surface Chemistry*

The surface chemistry of Ch was explored using XPS by comparing an untreated control with plasma-irradiated samples exposed to a fluence of $1 \times 10^{18}$ ions/cm$^2$ at irradiation angles of 45° and 60° to identify the surface functional groups introduced or modified by plasma treatment. In line with the multiscale PCA results (**Supplementary Figures S4 and S5**), these XPS measurements indicate that plasma irradiation induces systematic but comparatively subtle chemical changes that evolve consistently with the treatment-dependent surface restructuring.

**Figure 3a** illustrates the nitrogen chemical behavior as a function of argon plasma irradiation angle, with the observed broadening of the overall N1s spectrum indicating variations in chemical states relative to the untreated Ch. After extracting the binding energy positions through the curve-fitting of the N1s signal, its corresponding chemical states were identified (**Figure 3b-d**, Table 1). Therefore, the spectral features located at 399.4±0.2 eV [44,52–54] and 400.3±0.2 eV [44,52,55], appearing only for the control Ch, are attributed to C-$NH_2$ and C-N-C=O groups. At this stage, it is noteworthy that the 75.5% fraction of non-protonated amine observed in the control sample (Fig. 1b, Table 1) closely matches the nominal degree of acetylation specified for the commercial Ch [56]. Moreover, the quantitative analysis yields an atomic ratio of O to N of 4.5, and of C to O of 2.3, both slightly higher than the expected stoichiometric values of 4.0 and 1.5, respectively, for Ch, $[(C_6H_{11}NO_4)_n]$ [57].

The elevated C/O ratio presumably came from adventitious surface contamination. The nitrogen chemical response of Ch to the plasma treatment conditions was linked to the creation of two additional functional groups, namely N=C and $NH_3^+$ for the 45-degree irradiation and $NH_3^+$ and $NH_4^+$ for the 60-degree irradiation, with their corresponding binding energies located at 398.0±0.2 eV [58–61], 401.1±0.2 eV [52–54,56,62], and 402.6±0.2 eV [63–66], respectively. More specifically, plasma irradiation induced surface amination and protonation, resulting in the incorporation of $NH_4^+$ functionalities exclusively at higher irradiation angles (**Figure 5c-d, Table 1**). In terms of atomic fractions, the free amine content decreased by approximately half in both cases, dropping from ~75% to 34% and 39%, respectively (**Figure 5b-d**, Table 1). At the same time, the O/N ratio decreased, from 4.5 to 2.4 (**Table 2**).

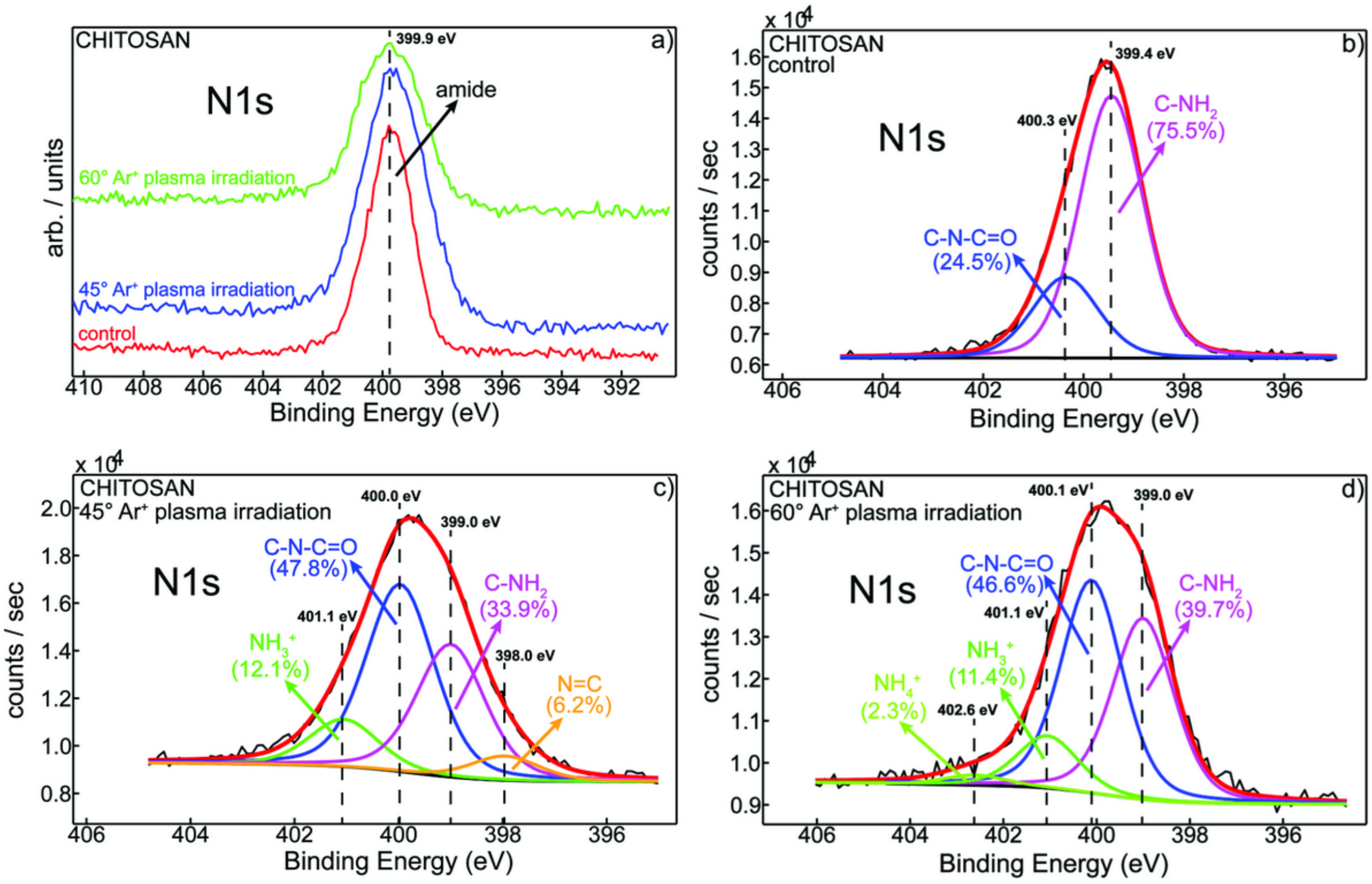

**Figure 3**. N1s XPS superimposed spectra of the Ch before and after $Ar^+$ plasma irradiation a); N1s XPS peak-fitted spectrum of the Ch before $Ar^+$ plasma irradiation b); N1s XPS peak-fitted spectrum of the Ch after 45° $Ar^+$ plasma irradiation c); N1s XPS peak-fitted spectrum of the Ch after 60° $Ar^+$ plasma irradiation d).

**Table 1. Nitrogen chemical table: chemical species, binding energies, and chemical state relative concentrations.**

| Ch | Nitrogen chemical species | Binding energy [eV] | Nitrogen chemical states relative conc. [%] |
|---|---|---|---|
| control | $C\text{-}NH_2$ | 399.4 | 75.5 |
| | C-N-C=O | 400.4 | 24.5 |
| 45° $Ar^+$ irradiation | N=C | 398.0 | 6.2 |
| | $C\text{-}NH_2$ | 399.0 | 33.9 |
| | C-N-C=O | 400.0 | 47.8 |
| | $NH_3^+$ | 401.1 | 12.1 |
| 60° $Ar^+$ irradiation | $C\text{-}NH_2$ | 399.0 | 39.7 |
| | C-N-C=O | 400.1 | 46.6 |
| | $NH_3^+$ | 401.1 | 11.4 |

| | $NH_4^+$ | 402.6 | 2.3 |
|---|---|---|---|

Further, **Figure 4a** presents the carbon spectra of Ch before and after plasma treatment, highlighting the general trend with an increasing irradiation angle. At first glance, the amine and amide groups show a visible decrease, accompanied by a pronounced increase in aliphatic carbon species. In addition, the former observation is consistent with the decreasing trend of nitrogen amines (**Figure 3**, **Table 1**).

For a more refined perspective, the C1s peak signal was decomposed into its components with binding energies of 284.9±0.2 eV [44,67,68], 286.4±0.2 eV [44,67,68], 288.0±0.2 eV [44,67,68], 289.5±0.2 eV [52,68], and 291.0±0.2 eV [52,69], characteristic of C-C/C-H, C-O/C-$NH_2$, C=O/C-N-C=O, O-C=O/C-$N_3$ bonds and shake-up satellite, respectively. Quantitatively, the fraction of non-protonated amines decreases with an increasing irradiation angle, nearly halving from 45.4% to 23.3% (**Figure 4b-d**, **Table 3**). This trend aligns well with the above nitrogen chemical behavior, which implies a similar reduction in the amine content (**Figure 3, Table 1**). On the other hand, the C-to-O ratio increases with the angle of irradiation (**Table 2**), accompanied by a decrease in the C-O chemical bond, which cannot be completely ruled out (**Figure 4b-d**, **Table 3**).

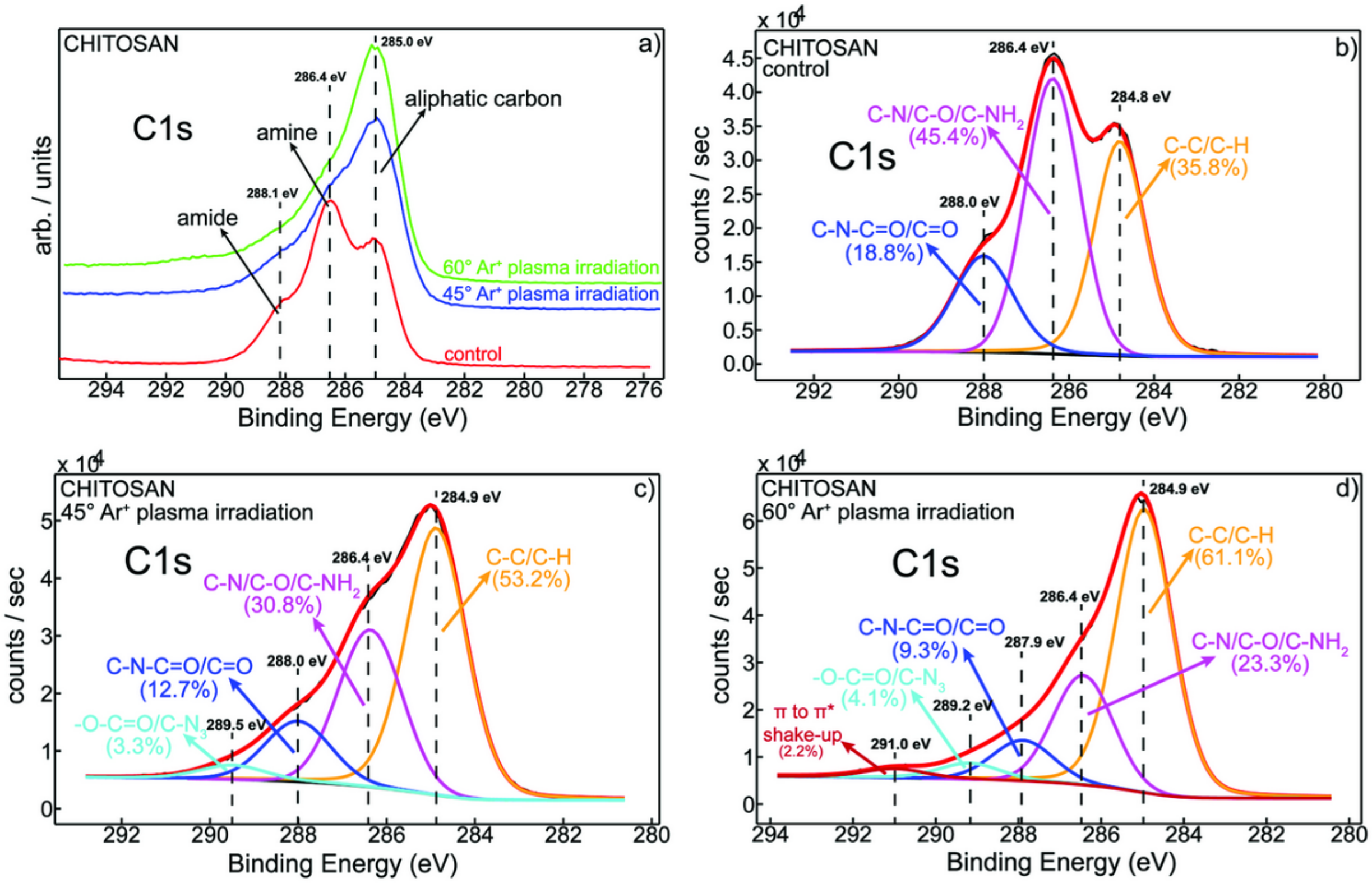

**Figure 4. C1s XPS superimposed spectra of the Ch before and after $Ar^+$ plasma irradiation a)**; C1s XPS peak-fitted spectrum of the Ch before $Ar^+$ plasma irradiation b); C1s XPS peak-fitted spectrum of the Ch after 45° $Ar^+$ plasma irradiation c); C1s XPS peak-fitted spectrum of the Ch after 60° $Ar^+$ plasma irradiation d).

**Table 2. Element relative concentrations (at. %).**

| Ch | C1s | N1s | O1s | C/O atomic ratio | O/N atomic ratio |
|---|---|---|---|---|---|
| control | 65.2±6.5 | 6.3±0.6 | 28.5±2.8 | 2.3 | 4.5 |
| 45° $Ar^+$ irradiation | 68.8±6.8 | 8.7±0.8 | 22.5±2.2 | 3.1 | 2.6 |
| 60° $Ar^+$ irradiation | 78.6±7.8 | 6.3±0.6 | 15.1±1.5 | 5.2 | 2.4 |

In terms of atomic balance, specifically the comparison between the total number of atoms and the fraction that are chemically bonded, we obtained the following results, derived from oxygen quantification using both elemental and carbon curve-fits, where the summed areas of the nitrided and oxidized carbon peaks, normalized to the total carbon signal should equal the ratio of oxygen and nitrogen concentrations to carbon [52,70], and only minor differences were observed:

Control: (6.3+28.5)/65.2=0.53; (45.4+18.8)/100=0.64

45° $Ar^+$ irradiation: (8.7+22.5)/68.8=0.45; (30.8+12.7+2*3.3)/100=0.50

60° $Ar^+$ irradiation: (6.3+15.1)/78.6=0.27; (23.3+9.3+2*4.1)/100=0.40

**Table 3. Carbon chemical table: chemical species, binding energies, and chemical state relative concentrations.**

| Ch | Carbon chemical species | Binding energy [eV] | Carbon chemical states relative concentrations [%] |
|---|---|---|---|
| control | C-C/C-H | 284.8 | 35.8 |
| | C-N/C-$NH_2$ | 286.4 | 45.4 |
| | C-N-C=O/C=O | 288.0 | 18.8 |
| 45° $Ar^+$ irradiation | C-C/C-H | 284.9 | 53.2 |
| | C-N/C-$NH_2$ | 286.4 | 30.8 |
| | C-N-C=O/C=O | 288.0 | 12.7 |
| | -O-C=O/C-$N_3$ | 289.5 | 3.3 |
| 60° $Ar^+$ irradiation | C-C/C-H | 284.9 | 61.1 |
| | C-N/C-$NH_2$ | 286.4 | 23.3 |
| | C-N-C=O/C=O | 287.9 | 9.3 |
| | -O-C=O/C-$N_3$ | 289.2 | 4.1 |
| | $\pi$-$\pi^*$ shake-up | 291.0 | 2.2 |

Since this is a follow-up study of our previous work [52], where we comprehensively examined the plasma-induced chemical changes in SF under identical experimental conditions, **Figure 5** presents only a direct comparison between Ch and SF as a function of irradiation angle. Therefore, a close inspection of the carbon signals reveals a completely different chemical profile for the control samples (**Figure 5a** and **b**). Specifically, Ch exhibits a clear enrichment in amine groups, while, at the same time, SF shows an increase in amide functionalities (**Figure 5a**), which is in turn consistent with the corresponding enhancement of nitrogen-bound amides in the latter (**Figure 5b**). Furthermore, the enrichment of amide groups in SF remains evident after 45° and 60° plasma irradiation, accompanied by a corresponding increase in amine groups in Ch (**Figure 5c,e**). From the nitrogen perspective, the amide contribution in SF increases after plasma treatment and becomes more pronounced at higher irradiation angles (**Figure 5d**,**f**). Simultaneously, its amine contribution reaches a higher level at 45° (**Figure 5d**), in agreement with the corresponding carbon chemistry (**Figure 5c**).

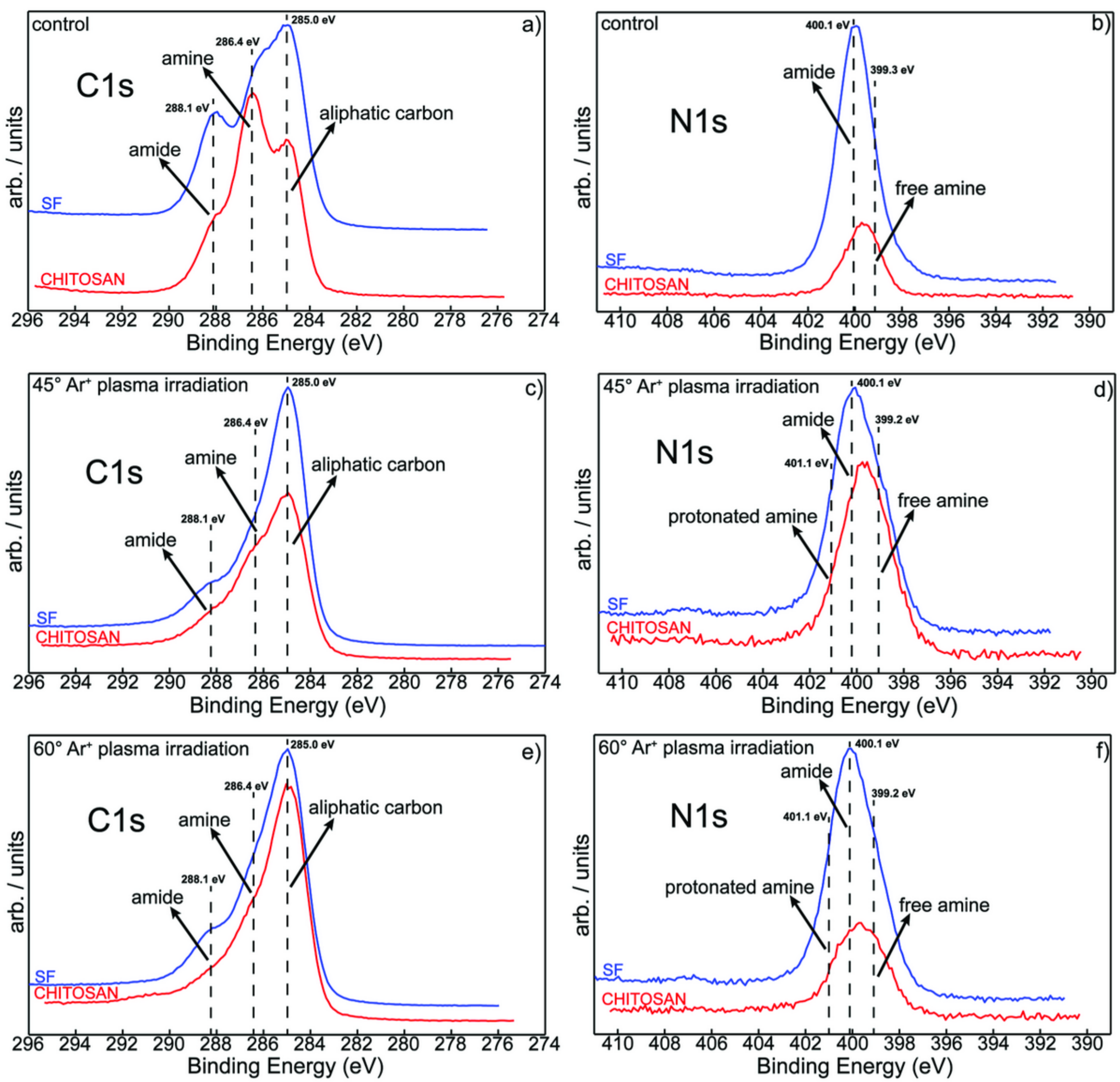

**Figure 5.** C1s XPS superimposed spectra for the Ch and SF: (a) control; (c) after 45° $Ar^+$ plasma irradiation; (e) after 60° $Ar^+$ plasma irradiation; N1s XPS superimposed spectra for the Ch and SF: (b) control; (d) after 45° $Ar^+$ plasma irradiation; (f) after 60° $Ar^+$ plasma irradiation.

Multiscale AFM analysis revealed that increasing the irradiation angle does not simply amplify surface roughness, but instead redistributes surface complexity, preferentially increasing curvature at smaller length scales and producing oriented nanofeatures with characteristic dimensions comparable to those of individual bacteria (~400 nm-1 μm).

Although XPS confirmed that plasma treatment induces measurable chemical changes, most notably surface amination and protonation in Ch and enhanced amide enrichment in SF, these modifications are systematic and relatively subtle compared with the pronounced topographic restructuring. Importantly, the chemical distinction between the two materials is preserved and modestly amplified by irradiation, such that Ch and SF exhibit distinct yet internally consistent combinations of surface chemistry and topographic texture at each treatment angle. This approach establishes a matrix of surfaces in which topographic complexity and scale distribution vary in a controlled, angle-dependent manner across two chemically distinct but comparably treated substrates, providing a robust experimental framework to elucidate how surface-feature scales, rather than bulk chemistry alone, govern biological responses.

Because wetting behavior is governed by the combined influence of surface chemistry and topography, future work could further examine how the interplay between these two factors influences wettability through direct contact-angle measurements. Such measurements would complement the present chemical and topographical characterization and help clarify how multiscale surface structure modulates interfacial interactions with biological systems.

### *3.2 Biological Response to DPNS-Treated Surfaces*

#### *3.2.1 Surface-Dependent Bacterial Attachment Behaviors from Nano- to Mesoscale*

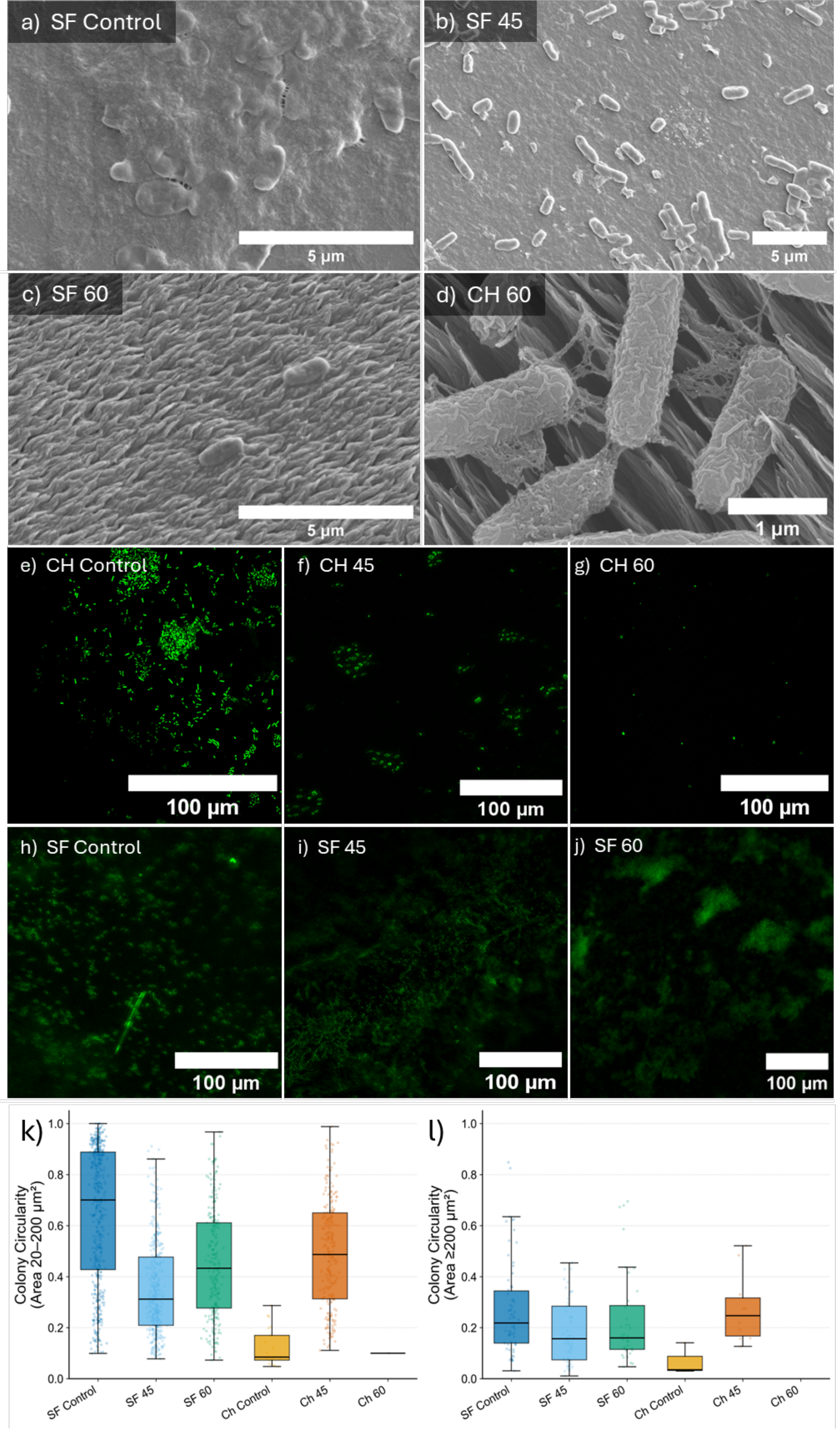

**Figure 6. Multiscale bacterial attachment on silk fibroin (SF) and chitosan (Ch) surfaces.** Representative SEM images showing bacterial adhesion on (a) SF Control, (b) SF 45, (c) SF 60, and (d) Ch 60. Panel (d) corresponds to a higher-magnification view highlighting local cell–surface interactions on Ch 60. Fluorescence images stained with SYTO 9 illustrate bacterial attachment patterns on (e) Ch Control, (f) Ch 45, (g) Ch 60, (h) SF Control, (i) SF 45, and (j) SF 60, capturing mesoscale organization over larger surface areas. (k) Colony circularity for medium-sized colonies (20–200 μm$^2$) and (l) large colonies (200 μm$^2$), where a value of 1.0 represents a perfectly circular morphology and lower values indicate irregular, dendritic, or spread biofilm architectures. Scale bars are indicated in each panel. Panels (d), (e), and (g) are adapted from [13] with permission. Copyright © 2023 American Chemical Society.

To compare attachment behaviors between the two materials and further describe the emerging multiscale correlation, we evaluated representative surface conditions for both SF and Ch. SF Control samples exhibited a well-developed and cohesive biofilm (**Figure 6a**). SF 45 showed more closely spaced bacteria (**Figure 6b**), whereas SF 60 displayed reduced aggregation (**Figure 6c**). In this context, SEM and fluorescence imaging (**Figure 6e-h**) probe different but complementary length scales. SEM captures nanoscale and microscale attachment features that are locally informative but spatially limited, whereas fluorescence imaging integrates these events over larger areas, enabling mesoscale quantification of area coverage and colony size. The combined use of both techniques therefore supports a multiscale interpretation of attachment behavior, rather than a one-to-one comparison between imaging modalities.

In contrast, Ch surfaces (previously characterized in [13]) revealed distinct attachment modes across their respective nanoscale features. Ch Control presented flat clusters without evidence of a mature biofilm. Samples with small nanostructures exhibited dispersed clusters with early polysaccharide-like material and occasional pili, while Ch-60 showed bacteria positioned between nanostructures, extending appendages across nanoscale features and displaying more prominent flagella (**Figure 6d**).

The structural organization of these colonies was further characterized by their circularity (**Figure 6k–l**). For medium colonies (20–200 μm$^2$), SF Control exhibited the highest median circularity (~0.7), which notably decreased following treatment (SF 45 and SF 60). This suggests that the features on irradiated SF drive more irregular or dendritic colony expansion compared to the pristine control. In contrast, Ch Control supported irregular colonies with the lowest circularity observed in the study (median $< 0.1$). However, Ch 45 supported more rounded clusters with higher circularity. In large colonies (≥ 200 μm$^2$), median circularity across all groups remained low (< 0.3), reflecting the inherent loss of symmetry as developing bacterial assemblies transition into complex, cohesive architectures.

These qualitative patterns align with the multiscale attachment mechanisms described for *E. coli*. While a detailed mechanistic description of bacterial attachment is beyond the scope of this work, the following observations are discussed in terms of potential correlations between multiscale surface descriptors and attachment patterns, supported by established literature. At the nanoscale to sub-microscale, pili and curli fimbriae adapt their length and orientation to engage with nanorough or patterned surfaces, tethering into nanoscale pits and ridges and promoting irreversible attachment [71]. Appendage length further modulates in response to encountered features (flagella elongate on nanorough alumina, pili extend in thin silica wells, and both shorten in confined geometries) supporting active adaptation to surface architecture [72,73]. This fine-scale stabilization complements body–surface realignment on nanostructured substrates [72], collectively reducing detachment probability and biasing cells toward early clustering.

These nanoscale anchoring events carry upward into the microscale regime, where flagella probe and hook into ridges, hollows, and hummocks [74]. When features provide mechanical "hooks," attachment persists; when they do not, reversible adhesion dominates and cells retain higher mobility [74]. Microscale valleys or depressions can also trap cells, limiting lateral encounters and delaying microcolony formation, depending on valley depth and peak-to-peak spacing [74].

These hierarchical mechanisms map directly onto our image-derived mesoscale metrics. Surfaces that, based on SEM, allowed closer packing or more continuous biofilm-like structures also showed increased area coverage and more frequent cluster formation (**Figure 6a** and **6e**). Conversely, surfaces with isolated or sparsely distributed bacteria corresponded to decreased coverage and fewer detected clusters (**Figure 6f** and **6h**). This agrees with the notion that once early attachment stabilizes, appendage entanglement and extracellular polymeric substances secretion (EPS) reinforce cell immobilization and multicellular cohesion [75], making mesoscale clustering the cumulative expression of nanoscale tethering and microscale confinement.

In addition to appendage-mediated attachment mechanisms, nanoscale surface features can also influence bacterial behavior through direct mechanical interactions at the cell–surface interface. Similar ion-induced nanostructured surfaces have been shown to induce membrane deformation and tension-driven responses, and in some cases membrane rupture, as a function of feature geometry and cell envelope properties [21]. In this context, the attachment patterns observed here may also reflect local mechanical interactions between bacterial cells and nanoscale surface features, which can modulate adhesion stability and subsequent spatial organization. However, as these early-stage interactions dictate the transition toward mature biofilm development, future studies should incorporate the quantification of EPS and evaluate bacterial responses in larger-scale, long-term cultures to fully resolve the impact of multiscale topography on biofilm stability and structural integrity.

#### *3.2.2 Area Coverage and Colony Size Distributions Across Treatments*

Area coverage measurements from SYTO 9 staining further support the qualitative trends and multiscale narrative (**Figure 7d**). SF 45 displayed the largest covered surface area, followed by SF 60. SF consistently exhibited much higher biofilm coverage than Ch (**Figure 7d**). Overall, within each material, the 45-degree samples exhibited the greatest area coverage, while the 60-degree samples decreased substantially toward control values, except for the noted increase in SF 60.

The size distribution for bacterial colonies was extremely right-skewed (**Supplementary Figure S2**), making summary statistics from the distribution challenging to interpret. Moreover, as discussed in the introduction, there is a strong scale-dependence to many observed surface-biology interactions [24,26,29]. Finally, individual bacteria themselves behave quite differently than biofilms [18,76], as isolated cells typically exhibit transient and reversible attachment, whereas clusters and biofilms represent progressively stabilized, multicellular assemblies whose formation reflects collective behavior across multiple length scales. These factors, combined with qualitative observations from SEM and fluorescence images, motivated the separation of bacterial structures by size. These size ranges were defined as 1-20 $\mu m^2$, 20-200 $\mu m^2$, and > 200 $\mu m^2$, shown in **Figure 7a-c**.

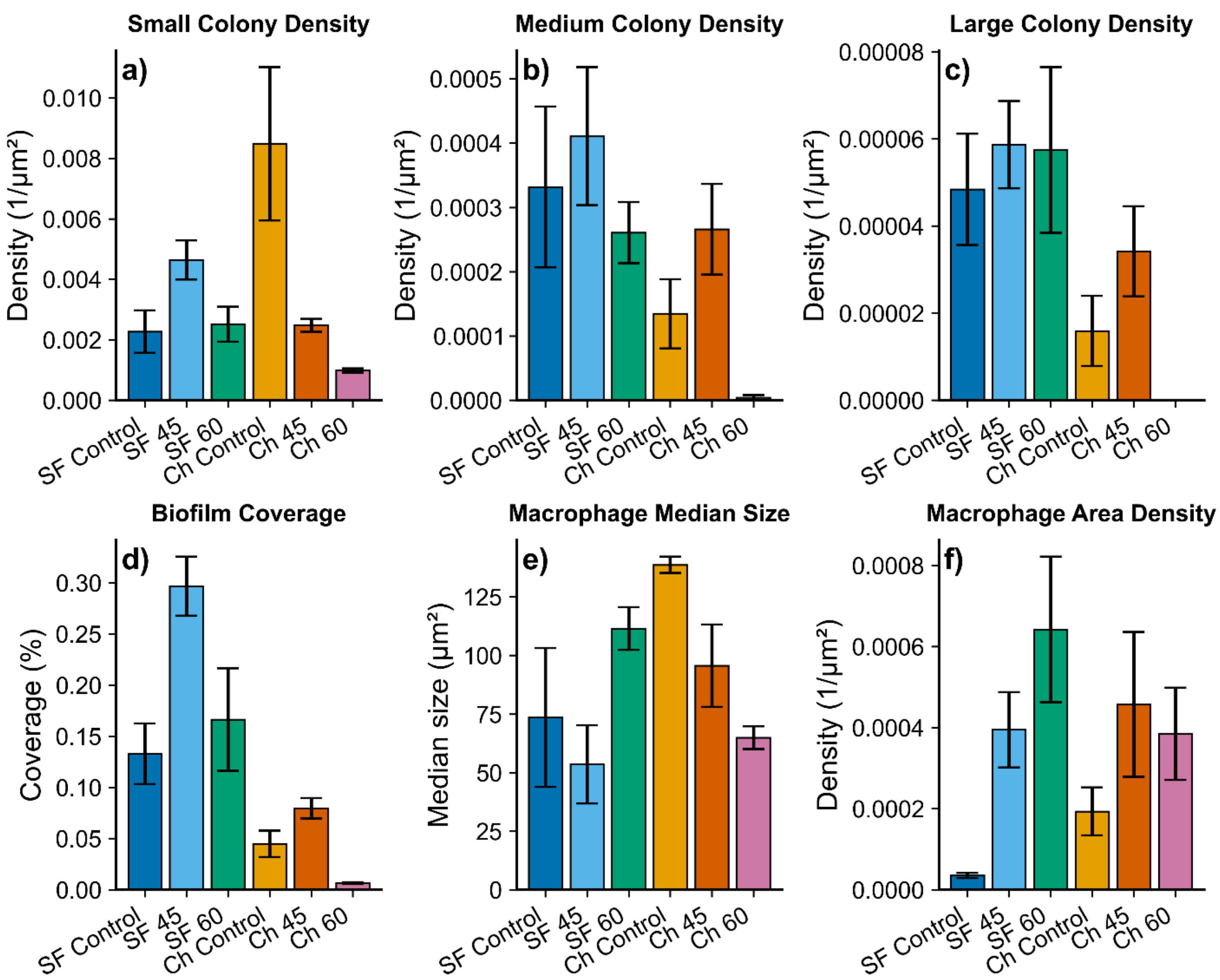

**Figure 7. Biological response data across the six material-treatment groups.** Each bar shown is the mean value across images along with standard error. Errors are descriptive, as images were not collected from independent samples. a) Area density of small (1-20 μm$^2$), b) medium (20-200 μm$^2$), and c) large (>200 μm$^2$) bacterial colonies, d) fractional biofilm area coverage, e) median macrophage size (area), and f) macrophage area density. All data is normalized by image area. Note that bacteria areal densities (a,b,c) are not on the same scale axes; small colonies are significantly more prevalent than medium and large ones (see **Supplementary Figure S2**).

Small (1-20 μm$^2$) colony density showed differing trends with respect to material and treatment; in the case of Ch films, small colony density continuously and sharply decreased from Ch Control to Ch 45, and to Ch 60, which exhibited the lowest number of small bacterial structures. SF showed a much different trend: density increased from SF Control to SF 45 and then decreased slightly to SF 60 back to near the control value. Thus, DPNS treatment seemed to elicit a larger change in small colony density in Ch than in SF. The six groups displayed some differences in their mitigation of medium bacterial colonies compared to small bacterial colonies. For SF, medium colony density trended similarly to small colony density. In the Ch samples, while Ch 60 remained the least supportive of bacterial growth, Ch 45 supported more intermediate-sized clusters than Ch Control. Interestingly, Ch Control supported fewer medium bacterial colonies than SF Control, despite supporting a much higher density of small colonies than SF Control, reflecting the qualitative observations of isolated clusters in Ch Control compared to the more continuous films in SF Control.

The apparent gaps between material and treatment widened further when observing the number of large bacterial colonies supported on the material surfaces. Large bacterial colony density was effectively unchanged from control to treated SF. Of the Ch groups, Ch 45 supported the highest density of large bacterial colonies while Ch 60 supported none. As with the medium colonies, Ch Control supported much fewer large bacterial colonies than SF Control. Collectively, the data indicated that treatment produced more marked changes in colony density in Ch across colony sizes than in SF. Treatment only seemed to impact the density of small bacteria in SF, and only marginally so. Moreover, Ch Control supported a high degree of isolated, small colonies, whereas SF Control was characterized instead by medium and large colonies.

Chitosan surfaces have been reported to be permissive to early bacterial contact, particularly when surface properties are modified in ways that favor adhesion. Under these conditions, chitosan's surface charge and mechanical compliance can enhance initial bacterial attachment, allowing bacteria to dock on the surface prior to subsequent stages such as biofilm maturation [77]. Importantly, early bacterial contact with a surface can be maintained even when progression toward mature biofilm formation is inhibited; a behavior often attributed to disrupted stabilization rather than reduced attachment frequency. In such cases, initial attachment may proceed normally, while later steps required for biofilm maturation, including extracellular polymeric substance production and multilayered cell accumulation, are impaired [78].

In contrast, multiple studies have reported that SF supports cohesive biofilm formation, even when initial bacterial attachment is comparatively limited. SF's moderate wettability and robust network structure, frequently associated with β-sheet-rich conformations, provide favorable conditions for bacterial adhesion and subsequent biofilm development. While initial attachment may still depend on specific surface properties and can be lower than on other substrates, once bacteria are present, SF promotes strong cell–cell interactions and biofilm cohesion, leading to the formation of mature and stable biofilm structures [79]. In this context, our results extend existing literature by demonstrating that these material-dependent behaviors emerge at distinct colony length scales: Ch surfaces exhibit a bottleneck between early attachment and cluster maturation that is highly sensitive to DPNS treatment, whereas SF surfaces maintain mesoscale organization that is comparatively robust to nanoscale modification.

#### *3.2.3 Macrophage Response*

To characterize macrophage adhesion behavior, we quantified the projected area of macrophage clusters formed on different surface conditions. Clusters were defined as spatially contiguous groups of adherent macrophages, representing localized regions of stabilized adhesion rather than isolated, highly motile cells. Macrophage attachment and clustering are mediated by dynamic cytoskeletal protrusions, including filopodia and lamellipodia, which actively probe the surface, establish adhesive contacts, and facilitate cell–cell interactions [80,81]. These appendage-like structures were qualitatively observed in SEM images as elongated protrusions extending from adherent macrophages and interacting with surface features. Unlike bacteria, whose dimensions are comparable to nanoscale surface features and can therefore be mechanically trapped or confined, macrophages operate at a much larger characteristic length scale (15-18 μm for J774 in spherical state [82]), passing over individual nanostructures and voids and instead integrating nanoscale cues through collective adhesion and spreading responses. Cluster area was therefore used as a mesoscale descriptor of collective cellular organization, integrating surface-mediated confinement and cell–cell proximity. While this analysis does not aim to infer macrophage activation state or fusion, it provides a quantitative framework to compare how multiscale surface properties influence the extent and organization of macrophage clustering.

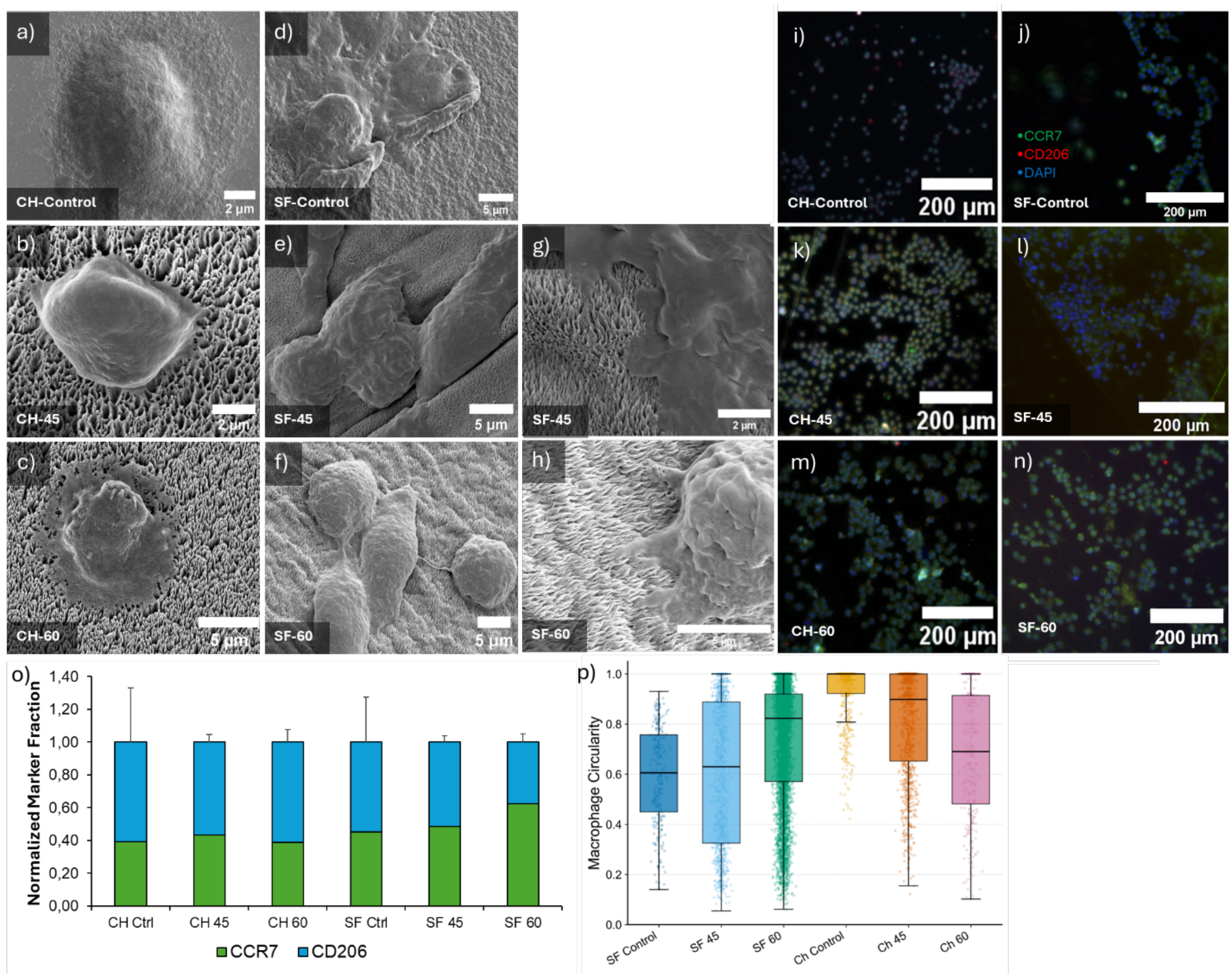

**Figure 8. Representative macrophage adhesion and morphology on chitosan (Ch) and silk fibroin (SF) surfaces under different surface treatments.** SEM images illustrate macrophage morphologies on chitosan surfaces: (a) Ch Control, (b) Ch 45, and (c) Ch 60, and on silk fibroin surfaces: (d) SF Control, (e) SF 45, and (f) SF 60. Higher-magnification SEM images of macrophages on SF surfaces are shown in (g) SF 45 and (h) SF 60, highlighting cell–surface interactions with the underlying topography. Fluorescence microscopy images show corresponding macrophage distributions on (i) Ch Control, (k) Ch 45, and (m) Ch 60, and on (j) SF Control, (l) SF 45, and (n) SF 60. CCR7-positive cells are shown in green, CD206-positive cells in red, and nuclei are counterstained with DAPI (blue). These images illustrate treatment-dependent differences in projected cell area, spatial distribution, and clustering behavior at the population level. (o) Quantification of normalized CCR7 and CD206 fractions, showing the relative contribution of each marker to total fluorescence intensity (CCR7 + CD206 = 1). p) Macrophage circularity across Ch and SF surfaces, where a value of 1.0 represents a perfectly circular

morphology and lower values indicate increased cell spreading and elongation. Panels (d–h) and (l, n) were adapted from [7] with permission.

SEM observations were used to resolve macrophage adhesion and spreading behaviors at the local scale. Across the Ch surfaces, macrophage morphology was modulated by treatment. On Ch Control, macrophages appeared extensively spread with large projected contact areas, reflecting stable cell–surface interactions. Ch 45 surfaces supported macrophages with more compact morphologies and reduced spreading, indicative of attachment events that remain weakly stabilized. In contrast, Ch 60 exhibited more flattened macrophages with larger contact areas than Ch 45, suggesting enhanced cell-surface adhesion.

Median macrophage particle size showed distinct trends with respect to material and treatment. In the Ch samples, the median macrophage size decreased continuously from Ch Control to Ch 45 to Ch 60. In contrast, the median macrophage size on SF was the lowest for the 45-degree sample and highest for the 60-degree sample. Macrophage area density increased sharply with respect to treatment in SF. Together, the median particle size and area density data showed that Ch Control was characterized by fewer, larger clusters relative to Ch 45 and Ch 60, which a higher density of smaller, isolated macrophages. The data for SF showed that SF control had a low area density of intermediate-sized particles, SF 45 supported intermediate macrophage density and size, and SF 60 had many large particles, indicating that irradiation of SF promoted macrophage attachment and spreading.

The behavior moves up a scale when individual cell actions result in population-level organization (mesoscale). Median macrophage particle size showed distinct trends with respect to material and treatment. In Ch samples, size decreased continuously from Ch Control to Ch 60. This was mirrored by circularity (**Figure 8p**), where Ch Control exhibited the highest median value (~0.95), indicating a predominantly rounded morphology, while Ch 60 showed increased heterogeneity. In contrast, SF size was lowest for SF 45 and highest for SF 60. Together, size, density, and circularity data (**Figure 8p**) showed that Ch Control featured fewer, larger clusters, while SF 60 supported many large particles with a broad range of circularity, indicating that irradiation of SF promoted enhanced attachment and spreading.

To complement these morphological observations, macrophage polarization was quantified through normalized CCR7 and CD206 fractions at 24 h (**Figure 8o**). In chitosan samples, CCR7/CD206 distributions remained relatively stable across treatments, whereas silk fibroin surfaces exhibited treatment-dependent shifts, with SF 60 showing an increased CCR7 fraction.

However, these differences were not statistically significant, indicating that the observed variations should be interpreted as trends rather than definitive phenotypic shifts.

At this early point, macrophage polarization is not fully established, and cells are expected to occupy intermediate activation states. Consistent with this, macrophages can co-express anti- and pro-inflammatory markers within the same population, reflecting a continuum of activation rather than discrete phenotypes [83]. Accordingly, the CCR7/CD206 fractions observed here represent early-stage macrophage responses that are predominantly governed by cell–material interactions, where cytokine signaling is present but not yet resolved into a stable phenotypic state.

### *3.3 Correlation of Biological Data with Surface Properties*

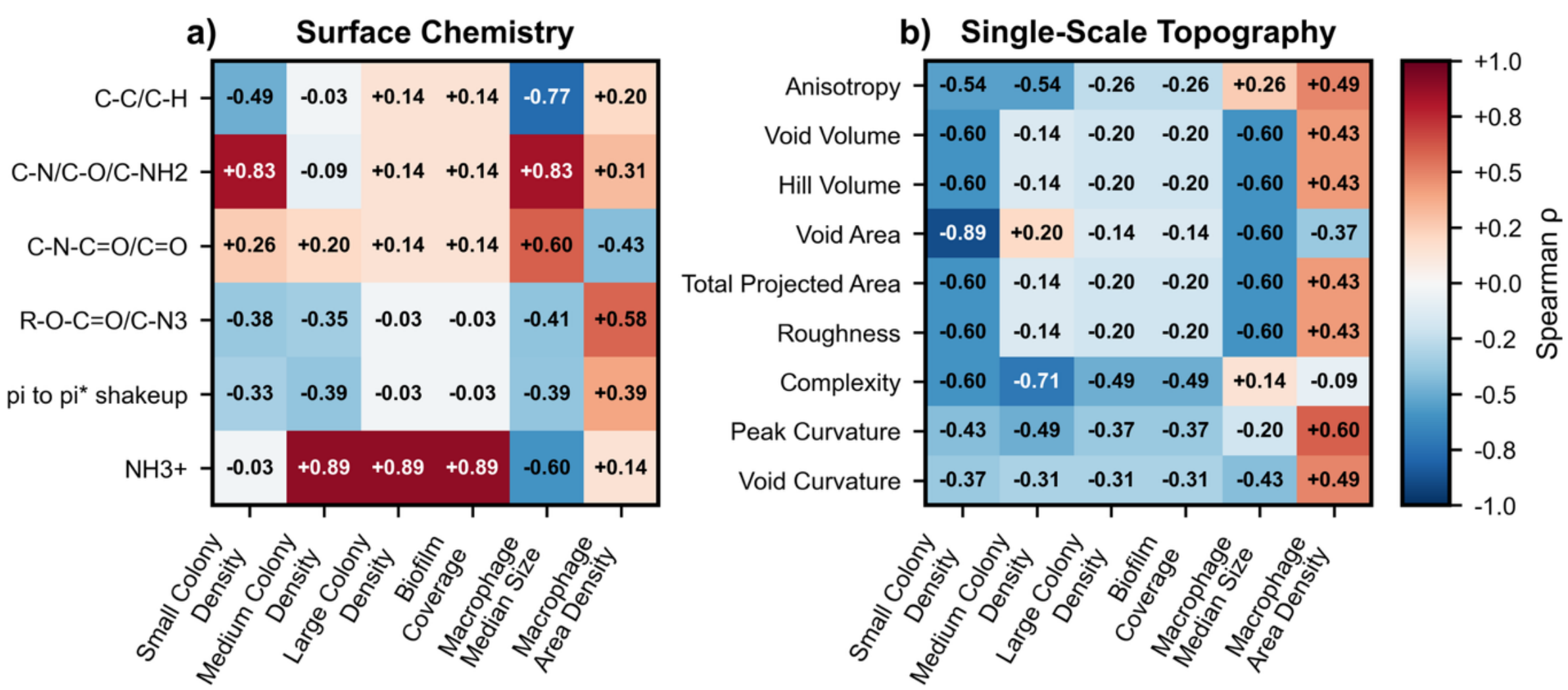

**Figure 9. Correlation heat maps of chemistry and single-scale topographic data with biological data.** a) surface chemistry and b) single-scale topography heat maps. Hill area was filtered out during significance testing and thus omitted from the single-scale heat map.

#### *3.3.1 Surface Chemistry:*

Although irradiation-induced chemical modifications are typically confined to the outermost nanometers of the surface [7], this region defines the biologically active interface governing protein adsorption and initial adhesion events [29,84]. Near-surface chemical functionalities regulate charge distribution, hydration, and short-range interactions, which collectively determine the composition and conformation of the adsorbed protein layer that mediates subsequent biological contact [85]. As a result, chemical variations restricted to the first few nanometers can influence collective biological organization, linking near-surface chemistry to multiscale adhesion behavior.

Data associated with larger bacterial colonies (large and medium colony density and biofilm coverage) showed a strong positive correlation with $NH_3^+$ concentration, whereas correlations with other functional groups were comparatively weak (**Figure 9a**). Because $NH_3^+$ levels were consistently higher on SF than on Ch and had low overall abundance (**Supplementary Figure S3**), this correlation may partially reflect material-specific chemistry rather than an isolated functional-group effect**.** Previous reports indicate that protonated amines confer a net positive charge to the near-surface region, which may favor electrostatic interactions with the negatively charged bacterial envelope characteristic of *E. coli* [86,87]. However, given the low overall abundance of these chemical groups (<3%, see **Supplementary Figure S3**), such interactions are more likely to contribute only modestly to attachment stabilization, especially when compared to the substantial influence of surface topography on bacterial adhesion [88].

In contrast, small bacterial colonies and isolated bacteria correlated positively with C–C/C–O/C–$NH_2$ concentration and showed a moderate negative relationship with C–C/C–H, but no evident association with protonated amines. These neutral or weak polar groups are typically linked to short-range interactions rather than long-range electrostatic forces, aligning with attachment behaviors that remain weakly stabilized and spatially limited [89,90].

J774 macrophages exhibited broader and stronger correlations with surface chemistry than bacteria, consistent with their mode of surface sensing. Unlike bacteria, which interact with substrates primarily through localized, appendage-mediated contacts, macrophages integrate interfacial cues over large contact areas through active spreading and cytoskeletal reorganization. It has been shown that J774 macrophages actively regulate their apparent surface area via cortical tension modulation and actin remodeling, enabling surface integration across micrometer-length scales [82,91]. This capacity for large-scale deformation enables macrophages to integrate chemical and topographic cues simultaneously, such that surface chemistry may exert a stronger influence when supported by favorable adhesion geometry rather than acting independently.

In this context, macrophage median size showed a high correlation with C–C/C–O/C–$NH_2$ concentration, a moderate positive correlation with C–N–C=O/C=O concentration, and negative correlations with C–C/C–H and $NH_3^+$ groups, suggesting an association between surface chemistry and macrophage spreading and projected contact area, rather than simple attachment. Moreover, macrophage area density exhibited a moderate positive correlation with R–O–C=O/C–

$N_3$ functionalities, suggesting chemistry-dependent differences in cell clustering. Neutral or moderately polar surface chemistries have been shown to promote selective adsorption and preserved bioactivity of serum proteins, thereby defining the biological identity of the interface and regulating macrophage interactions through integrin-mediated pathways [92].

### 3.3.2 *Single-Scale Topography*

Single scale topography correlations (**Figure 9b**) mirrored general results of the chemistry data. Broadly speaking, small bacterial structures and macrophages again showed strong correlation over a broad range of parameters, indicating a more robust link with surface properties and, by extension, material/treatment. Topographic parameters associated with roughness, complexity, volume, and area were closely related, as indicated by the similar correlation strengths of those families. Medium bacterial colony density showed correlation with only anisotropy and peak curvature while large bacterial colonies and area coverage showed weak correlation with all topographic parameters, indicating decreasing correlation strength with increasing colony size. Macrophage median size again showed stronger overall correlation with surface properties than macrophage area density, exhibiting a moderate negative correlation across volume, area, and roughness-associated parameters. Macrophage area density, on the other hand, correlated more strongly with complexity, curvature, and anisotropy.

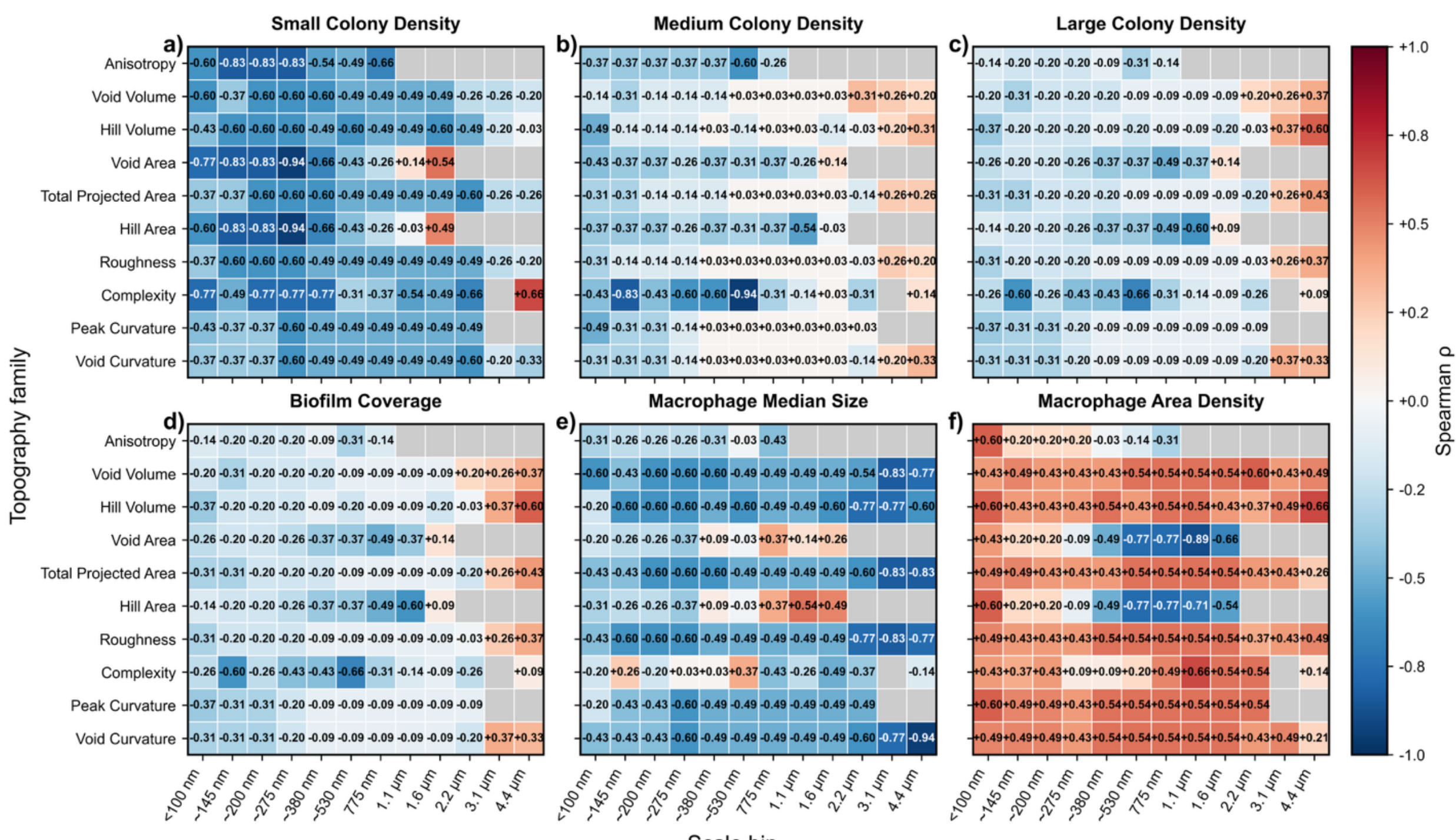

**Figure 10: Correlation heat maps of multiscale topographic data with biological data.** a) Small colony density, b) medium colony density, c) large colony density, d) biofilm coverage, e) median macrophage size, and f) macrophage area density. Plots show scale-dependence of correlation strength with respect to biological entity.

### *3.3.3 Multiscale Topography: Bandpass Data*

Multiscale topography data offered deeper insight into the scale-dependence of correlations in the biological data. Interestingly, broad relationships between the scale of biological features and the scale of observation which showed the strongest correlation could be observed. Small colony density showed the strongest and broadest correlation with bandpass data, including parameters associated with volume, area, roughness, anisotropy, and complexity, reinforcing the single-scale results. Importantly, bandpass data showed an increase in correlation strength at smaller scales, near 145-380 nm, and a strong decrease in correlation strength at larger scales, suggesting a scale-dependent relationship between surface topography and small colony density.

Larger bacterial structures generally showed weaker correlation with topographic parameters. Importantly, the scale of maximum correlation was roughly 500 nm – 1.1 µm, slightly larger than that for small bacterial colonies, and was stronger for medium colony density than for large colony density, again reflecting a decrease in correlation between colony response and surface conditions with increasing colony size. The density of medium and large structures was notably higher in SF than in Ch, irrespective of treatment.

These data indicate a complex, size- and material-dependent relationship wherein topography alone has decreasing correlation strength with increasing colony size. Topography exerted a weakening impact on larger bacterial structures in SF, whereas treated Ch mitigated these structures readily, and with increasing ability as topography became more pronounced. Thus, the efficacy of the tested topography in mitigating biofilm formation is likely contingent on both the size of the colony as well as intrinsic material properties. Given the generally weak correlation of large colonies with the collected data, it is likely that testing combination effects or higher-level properties (e.g., wettability, surface charge) is necessary to accurately assess material performance in this regard.

The data suggest that curvature and complexity near the characteristic scale (width) of individual *E. coli* (0.4-1 µm, [25]) limit biofilm formation and attachment, which could occur through a variety of mechanisms. Nanostructures may limit mobility and prevent interconnection via pili, potentially inducing stress [93] and limiting biofilm formation. Similarly, an “ordering” effect [88] could be manifested by bacteria preferentially occupying valleys between nanostructures to maximize contact area [71]. These mechanisms differ from bactericidal nanotopographies, which typically rely on smaller, higher aspect-ratio structures that induce mechanical deformation of the outer membrane or peptidoglycan wall of Gram-negative or Gram-positive bacteria, respectively [19,25,88]. A more limited mechanism could involve inhibition of initial attachment by preventing flagellar attachment to the substrate by tall nanostructures; however, this is unlikely to be the dominant effect, as such structures must be ~10 µm or taller [24], and treated SF clearly shows strong bacterial attachment.

Given the observed differences in material behavior, the ability to prevent biofilm formation via the generated topography may be contingent on additional biological stressors. At physiological pH, chelation of metal cations present in the bacterial cell membrane by amino groups is the dominant reported antibacterial mechanism of Ch [94], which is supported by the chemical correlations. Such chelation has been suggested to induce membrane stress and leakage [9], with conflicting data regarding differential impact on bacterial strain [94,95]. In this context, nanoscale topography may act primarily by trapping or segregating bacteria and limiting cooperative biofilm organization, while chemical effects provide an additional source of stress.

While the bacterial response appears largely constrained by geometric interactions at the scale of individual cells, the response of larger eukaryotic cells reflects a different mode of interaction with surface topography. To explore these effects, macrophage behavior was analyzed in relation to the same multiscale surface descriptors. Macrophage data correlated strongly with both multiscale and single-scale topographic descriptors. Macrophage median size exhibited broad negative correlation with volume, area, and roughness and strong negative correlation with void curvature. Notably, these correlation strengths were highest at the absolute largest scales of measurement, from 2.2–4.1 µm.

We hypothesize that biological sensitivity to surface curvature is governed by dimensional matching between operative cellular structural elements and surface feature size within the local topographic window sampled by AFM and the sensitive scales identified by multiscale analysis. When curvature within this window approaches the characteristic scale of bacterial width (~0.4–1 µm), geometric interference may reduce stable attachment and hinder cooperative biofilm assembly [15]. In contrast, for larger eukaryotic cells, curvature at nanoscale-to-micron scales may modulate adhesion indirectly through protein-mediated integrin engagement and focal adhesion maturation, which occur locally and are integrated across the cell footprint during spreading [96].

It is important to distinguish between the projected diameter of fully spread J774 macrophages (~15–20 μm) [82] and the characteristic length of scales at which surface sensing occurs. Macrophage adhesion and spreading are mediated by localized protrusive structures and focal adhesion complexes operating at submicron-to-micron dimensions. Filopodia exhibits diameters of ~100–300 nm and lengths of 1–10 μm [97], while lamellipodial protrusion is driven by actin polymerization at the leading edge [91]. These lamellipodial extensions span several microns laterally, with thicknesses on the order of ~100–200 nm and engage the substrate through discrete focal adhesion complexes measuring ~0.5–5 μm in length and ~0.2–1 μm in width [98].

The 5 × 5 μm AFM window therefore characterizes local curvature and roughness gradients encountered during adhesion and cytoskeletal organization rather than representing the entire projected cell footprint. Macrophages integrate multiple such local regions across the substrate during spreading.

Importantly, macrophage spreading is mediated through integrin-based focal adhesion complexes that couple extracellular ligand engagement to actin cytoskeletal organization and intracellular force transmission [99]. Nanoscale curvature can influence the spatial organization and conformation of adsorbed adhesion-mediating proteins [100,101], thereby modulating integrin clustering and focal adhesion maturation [102]. Because focal adhesion maturation regulates actin stress fiber formation and cytoskeletal tension [99], multiscale topographic features may propagate mechanical cues from the cell–substrate interface to the nucleus. While direct assessment of mechanotransduction pathways such as YAP/TAZ signaling was beyond the scope of this study, the observed scale-dependent correlations are consistent with curvature-mediated modulation of adhesion stabilization and cytoskeletal organization.

## 4. Conclusions and Outlook

Multiscale analysis was used to investigate the surface properties of DPNS-treated SF and Ch films and their relationship with macrophage and bacteria behavior. DPNS produced multiscale topographic alterations to both Ch and SF along with surface carbonization, deoxygenation, and changes in nitrogen chemistry, with topographic changes being more pronounced in Ch than SF. Biofilm formation was substantially reduced on treated Ch relative to SF, indicating material-dependent differences in the response of large bacterial structures to surface modification. This further suggests combinatorial approaches, as opposed to topography alone, are required to mitigate biofilm formation.

Multiscale data revealed scale-dependence in biological interactions with surface topography; individual bacteria and small colonies were strongly negatively correlated with small-scale topography, whereas correlation strength weakened as bacterial colony size increased. In contrast, macrophage response correlated most strongly with large-scale surface features, reflecting their larger size and integrated adhesion and spreading behavior. Opposing trends in cluster size and area density across scale indicate that surface features regulate both macrophage spreading and spatial organization. The above clearly demonstrates that scale is a relevant factor in understanding the role of topography in impacting biological response.

The introduction of two multiscale analysis techniques here provides a framework for scale-aware, quantitative analysis of complex biomaterial surfaces. Bandpass filtration captured a wider variety of topographic parameters than the multiscale curvature tensor method and thus was generally better able to capture correlations with biological response. Though single-scale methods could characterize key topographic features and associated correlations, multiscale methods demonstrated stronger correlations, were better able to discriminate surfaces, and, most importantly, could ascribe to these specific scales, enabling a more detailed analysis of the material performance and the impact of DPNS treatment.

A limitation in the present work was limited sample size compared to the parameter space (i.e., $n << p$). Further studies should collect more data to improve statistical power, particularly for multiscale data, allowing direct interpretation of the key surface features and scales driving biological response. Due to the extreme multicollinearity in multiscale data, efforts to cluster or prune data prior to analysis are recommended. Partial Least Squares (PLS) regression is a viable statistical framework to infer causality in this context if sample size is sufficiently large [103–105]. Additionally, the AFM scan size (5 × 5 μm) does not capture larger-scale surface gradients comparable to the full macrophage footprint. Although the selected window enables high-resolution multiscale curvature analysis, future work incorporating larger-area topographic mapping would further connect multiscale surface structure with whole-cell morphology. Finally, broadening the set of characterization parameters to include surface charge and wettability and/or allowing interaction between factors in statistical modelling would allow more comprehensive characterization data and analysis. Given the importance of protein adhesion in dictating subsequent cell attachment and behavior, direct investigation of protein adhesion via multiscale analysis may provide more direct mechanistic connections between surface properties and performance.

## 5. Acknowledgement

The authors acknowledge the Materials Research Institute (MRI) at Penn State for access to shared instrumentation, as well as technical training and expert support, including use of the Thermo Scientific Verios G4 scanning electron microscope (SEM), Thermo Scientific Scios FIB-SEM, Physical Electronics VersaProbe III X-ray photoelectron spectrometer (XPS), and Bruker Icon Dimension II atomic force microscope (AFM).

### 5.1.1 Declaration of generative AI and AI-assisted technologies in the manuscript preparation process

During the preparation of this work, the authors used artificial intelligence-assisted tools, including ChatGPT, Perplexity, and Gemini in order to help write code for parsing, analyzing, and plotting data. After using this tool/service, the authors reviewed and edited the content as needed and take full responsibility for the content of this published article.

## 6. Supporting Information

Supporting information: Single-scale topographic data, bacterial colony size distributions, functional group abundances, PCA score plots and loadings, multiscale curvature tensor heat maps, multiscale data relative scores, and topographic family designations and ISO definitions (DOCX).

## 7. References

[1] Percival SL, Hill KE, Williams DW, Hooper SJ, Thomas DW, Costerton JW. A review of the scientific evidence for biofilms in wounds. Wound Repair and Regeneration 2012;20:647–57. https://doi.org/10.1111/j.1524-475X.2012.00836.x.

[2] Wolcott R d., Rhoads D d., Dowd S e. Biofilms and chronic wound inflammation. J Wound Care 2008;17:333–41. https://doi.org/10.12968/jowc.2008.17.8.30796.

[3] dos Santos FV, Siqueira RL, de Morais Ramos L, Yoshioka SA, Branciforti MC, Correa DS. Silk fibroin-derived electrospun materials for biomedical applications: A review. International Journal of Biological Macromolecules 2024;254:127641. https://doi.org/10.1016/j.ijbiomac.2023.127641.

[4] Fazal T, Murtaza BN, Shah M, Iqbal S, Rehman M, Jaber F, et al. Recent developments in natural biopolymer based drug delivery systems. RSC Adv n.d.;13:23087–121. https://doi.org/10.1039/d3ra03369d.

[5] Ghalei S, Handa H. A review on antibacterial silk fibroin-based biomaterials: current state and prospects. Materials Today Chemistry 2022;23:100673. https://doi.org/10.1016/j.mtchem.2021.100673.

[6] González-Restrepo D, Zuluaga-Vélez A, Orozco LM, Sepúlveda-Arias JC. Silk fibroin-based dressings with antibacterial and anti-inflammatory properties. European Journal of Pharmaceutical Sciences 2024;195:106710. https://doi.org/10.1016/j.ejps.2024.106710.

[7] Posada VM, Marin A, Mesa-Restrepo A, Nashed J, Allain JP. Enhancing silk fibroin structures and applications through angle-dependent Ar+ plasma treatment. International Journal of Biological Macromolecules 2024;257:128352. https://doi.org/10.1016/j.ijbiomac.2023.128352.

[8] Qi Y, Wang H, Wei K, Yang Y, Zheng R-Y, Kim IS, et al. A Review of Structure Construction of Silk Fibroin Biomaterials from Single Structures to Multi-Level Structures. Int J Mol Sci 2017;18:237. https://doi.org/10.3390/ijms18030237.

[9] Yan D, Li Y, Liu Y, Li N, Zhang X, Yan C. Antimicrobial Properties of Chitosan and Chitosan Derivatives in the Treatment of Enteric Infections. Molecules 2021;26:7136. https://doi.org/10.3390/molecules26237136.

[10] Liu H, Wang C, Li C, Qin Y, Wang Z, Yang F, et al. A functional chitosan-based hydrogel as a wound dressing and drug delivery system in the treatment of wound healing. RSC Adv 2018;8:7533–49. https://doi.org/10.1039/C7RA13510F.

[11] Matica MA, Aachmann FL, Tøndervik A, Sletta H, Ostafe V. Chitosan as a Wound Dressing Starting Material: Antimicrobial Properties and Mode of Action. International Journal of Molecular Sciences 2019;20:5889. https://doi.org/10.3390/ijms20235889.

[12] Silva SS, Luna SM, Gomes ME, Benesch J, Pashkuleva I, Mano JF, et al. Plasma Surface Modification of Chitosan Membranes: Characterization and Preliminary Cell Response Studies. Macromolecular Bioscience 2008;8:568–76. https://doi.org/10.1002/mabi.200700264.

[13] Jaramillo-Correa C, Posada VM, Nashed J, Civantos A, Allain JP. Analysis of Antibacterial Efficacy and Cellular Alignment Regulation on Plasma Nanotextured Chitosan Surfaces. Langmuir 2023;39:14573–85. https://doi.org/10.1021/acs.langmuir.3c01808.

[14] Bandara CD, Singh S, Afara IO, Wolff A, Tesfamichael T, Ostrikov K, et al. Bactericidal Effects of Natural Nanotopography of Dragonfly Wing on Escherichia coli. ACS Appl Mater Interfaces 2017;9:6746–60. https://doi.org/10.1021/acsami.6b13666.

[15] Ivanova EP, Hasan J, Webb HK, Truong VK, Watson GS, Watson JA, et al. Natural bactericidal surfaces: mechanical rupture of Pseudomonas aeruginosa cells by cicada wings. Small 2012;8:2489–94. https://doi.org/10.1002/smll.201200528.

[16] Watson GS, Green DW, Schwarzkopf L, Li X, Cribb BW, Myhra S, et al. A gecko skin micro/nano structure – A low adhesion, superhydrophobic, anti-wetting, self-cleaning, biocompatible, antibacterial surface. Acta Biomaterialia 2015;21:109–22. https://doi.org/10.1016/j.actbio.2015.03.007.

[17] Linklater DP, Juodkazis S, Rubanov S, Ivanova EP. Comment on “Bactericidal Effects of Natural Nanotopography of Dragonfly Wing on Escherichia coli.” ACS Appl Mater Interfaces 2017;9:29387–93. https://doi.org/10.1021/acsami.7b05707.

[18] Khatoon Z, McTiernan CD, Suuronen EJ, Mah T-F, Alarcon EI. Bacterial biofilm formation on implantable devices and approaches to its treatment and prevention. Heliyon 2018;4:e01067. https://doi.org/10.1016/j.heliyon.2018.e01067.

[19] Modaresifar K, Azizian S, Ganjian M, Fratila-Apachitei LE, Zadpoor AA. Bactericidal effects of nanopatterns: A systematic review. Acta Biomaterialia 2019;83:29–36. https://doi.org/10.1016/j.actbio.2018.09.059.

[20] Arias SL, Cheng MK, Civantos A, Devorkin J, Jaramillo C, Allain JP. Ion-Induced Nanopatterning of Bacterial Cellulose Hydrogels for Biosensing and Anti-Biofouling Interfaces. ACS Appl Nano Mater 2020;3:6719–28. https://doi.org/10.1021/acsanm.0c01151.

[21] Arias SL, Devorkin J, Spear JC, Civantos A, Allain JP. Bacterial Envelope Damage Inflicted by Bioinspired Nanostructures Grown in a Hydrogel. ACS Appl Bio Mater 2020;3:7974–88. https://doi.org/10.1021/acsabm.0c01076.

[22] Mesa-Restrepo A, Byers E, Brown JL, Ramirez J, Allain JP, Posada VM. Osteointegration of Ti Bone Implants: A Study on How Surface Parameters Control the Foreign Body Response. ACS Biomater Sci Eng 2024;10:4662–81. https://doi.org/10.1021/acsbiomaterials.4c00114.

[23] Anselme K, Davidson P, Popa AM, Giazzon M, Liley M, Ploux L. The interaction of cells and bacteria with surfaces structured at the nanometre scale. Acta Biomaterialia 2010;6:3824–46. https://doi.org/10.1016/j.actbio.2010.04.001.

[24] Gu H, Chen A, Song X, Brasch ME, Henderson JH, Ren D. How Escherichia coli lands and forms cell clusters on a surface: A new role of surface topography. Scientific Reports 2016;6. https://doi.org/10.1038/srep29516.

[25] Mirzaali MJ, van Dongen ICP, Tümer N, Weinans H, Yavari SA, Zadpoor AA. In-silico quest for bactericidal but non-cytotoxic nanopatterns. Nanotechnology 2018;29:43LT02. https://doi.org/10.1088/1361-6528/aad9bf.

[26] Wang Y, Subbiahdoss G, Swartjes J, van der Mei HC, Busscher HJ, Libera M. Length-Scale Mediated Differential Adhesion of Mammalian Cells and Microbes. Advanced Functional Materials 2011;21:3916–23. https://doi.org/10.1002/adfm.201100659.

[27] Franz S, Rammelt S, Scharnweber D, Simon JC. Immune responses to implants – A review of the implications for the design of immunomodulatory biomaterials. Biomaterials 2011;32:6692–709. https://doi.org/10.1016/j.biomaterials.2011.05.078.

[28] Luo J, Walker M, Xiao Y, Donnelly H, Dalby MJ, Salmeron-Sanchez M. The influence of nanotopography on cell behaviour through interactions with the extracellular matrix – A review. Bioactive Materials 2022;15:145–59. https://doi.org/10.1016/j.bioactmat.2021.11.024.

[29] Lord MS, Foss M, Besenbacher F. Influence of nanoscale surface topography on protein adsorption and cellular response. Nano Today 2010;5:66–78. https://doi.org/10.1016/j.nantod.2010.01.001.

[30] Brown CA, Hansen HN, Jiang XJ, Blateyron F, Berglund J, Senin N, et al. Multiscale analyses and characterizations of surface topographies. CIRP Annals 2018;67:839–62. https://doi.org/10.1016/j.cirp.2018.06.001.

[31] Bartkowiak T, Peta K, Królczyk J, Niesłony P, Bogdan-Chudy M, Przeszłowski Ł, et al. Wetting properties of polymer additively manufactured surfaces – Multiscale and multi-technique study into the surface-measurement-function interactions. Tribology International 2024;202:110394. https://doi.org/10.1016/j.triboint.2024.110394.

[32] Chauvy PF, Madore C, Landolt D. Variable length scale analysis of surface topography: characterization of titanium surfaces for biomedical applications. Surface and Coatings Technology 1998;110:48–56. https://doi.org/10.1016/S0257-8972(98)00608-2.

[33] Dumas V, Guignandon A, Vico L, Mauclair C, Zapata X, Linossier MT, et al. Femtosecond laser nano/micro patterning of titanium influences mesenchymal stem cell adhesion and commitment. Biomed Mater 2015;10:055002. https://doi.org/10.1088/1748-6041/10/5/055002.

[34] Gambardella A, Marchiori G, Maglio M, Russo A, Rossi C, Visani A, et al. Determination of the Spatial Anisotropy of the Surface MicroStructures of Different Implant Materials: An Atomic Force Microscopy Study. Materials 2021;14:4803. https://doi.org/10.3390/ma14174803.

[35] Watari S, Hayashi K, Wood JA, Russell P, Nealey PF, Murphy CJ, et al. Modulation of osteogenic differentiation in hMSCs cells by submicron topographically-patterned ridges and grooves. Biomaterials 2012;33:128–36. https://doi.org/10.1016/j.biomaterials.2011.09.058.

[36] Zink C, Hall H, Brunette DM, Spencer ND. Orthogonal nanometer-micrometer roughness gradients probe morphological influences on cell behavior. Biomaterials 2012;33:8055–61. https://doi.org/10.1016/j.biomaterials.2012.07.037.

[37] Peta K, Bartkowiak T, Rybicki M, Galek P, Mendak M, Wieczorowski M, et al. Scale-dependent wetting behavior of bioinspired lubricants on electrical discharge machined Ti6Al4V surfaces. Tribology International 2024;194:109562. https://doi.org/10.1016/j.triboint.2024.109562.

[38] Bartkowiak T, Berglund J, Brown CA. Multiscale Characterizations of Surface Anisotropies. Materials (Basel) 2020;13:3028. https://doi.org/10.3390/ma13133028.

[39] Berglund J, Agunwamba C, Powers B, Brown CA, Rosén B -G. On discovering relevant scales in surface roughness measurement—an evaluation of a band-pass method. Scanning 2010;32:244–9. https://doi.org/10.1002/sca.20168.

[40] Brown CA, Blateyron F, Berglund J, Murrison AJ, Jeswiet JJ. Spatial frequency decomposition with bandpass filters for multiscale analyses and functional correlations. Surf Topogr: Metrol Prop 2024;12:035031. https://doi.org/10.1088/2051-672X/ad6f2f.

[41] Rockwood DN, Preda RC, Yücel T, Wang X, Lovett ML, Kaplan DL. Materials fabrication from Bombyx mori silk fibroin. Nat Protoc 2011;6:1612–31. https://doi.org/10.1038/nprot.2011.379.

[42] John F. Hanlon, Timothy A. Gessert. Materials in vacuum. A users guide to vacuum technology, John Wiley & Sons, Ltd; 2023, p. 259–83. https://doi.org/10.1002/9781394174232.ch15.

[43] Schamis H, Jaramillo-Correa C, Parsons MS, Marchhart T, Allain JP, Hargrove C, et al. The ion-gas-neutral interactions with surfaces-2 (IGNIS-2) facility for the study of plasma–material interactions. Review of Scientific Instruments 2024;95:043508. https://doi.org/10.1063/5.0165857.

[44] George GA. High resolution XPS of organic polymers—the scienta ESCA 300 data base. G. Beamson and D. Briggs. John Wiley & Sons, Ltd, Chichester, 1992. Pp. 295, price £65.00. ISBN 0-471-93592-1. Polymer International 1994;33:439–40. https://doi.org/10.1002/pi.1994.210330424.

[45] Ghosh P, Mondal J, Ben-Jacob E, Levine H. Mechanically-driven phase separation in a growing bacterial colony. Proceedings of the National Academy of Sciences 2015;112:E2166–73. https://doi.org/10.1073/pnas.1504948112.

[46] Berkmans, F., Bartkowiak, T., Grochalski, K., Wieczorowski, M., & Bigerelle, M. Curvature-Based Multiscale Feature Extraction for Surface Quality Inspection in Manufacturing. The International Journal of Advanced Manufacturing Technology 2025. https://doi.org/10.1007/s00170-025-17145-8.

[47] Digital Surf. Mountains software 2025.

[48] ISO 25178-2:2012. Geometrical product specifications (GPS) — Surface texture: Areal — Part 2: Terms, definitions and surface texture parameters 2012.

[49] Sung CH (Joseph), Hao T, Fang H, Nguyen AT, Perricone V, Yu H, et al. Biological and Biologically Inspired Functional Nanostructures: Insights into Structural, Optical, Thermal, and Sensing Applications. Advanced Materials n.d.;n/a:e09281. https://doi.org/10.1002/adma.202509281.

[50] Xu X, Jia J, Guo M. The Most Recent Advances in the Application of Nano-Structures/Nano-Materials for Single-Cell Sampling. Front Chem 2020;8. https://doi.org/10.3389/fchem.2020.00718.

[51] Gogolides E, Constantoudis V, Kokkoris G, Kontziampasis D, Tsougeni K, Boulousis G, et al. Controlling roughness: from etching to nanotexturing and plasma-directed organization on organic and inorganic materials. J Phys D: Appl Phys 2011;44:174021. https://doi.org/10.1088/0022-3727/44/17/174021.

[52] Posada VM, Ramírez J, Civantos A, Fernández-Morales P, Allain JP. Ion-bombardment-driven surface modification of porous magnesium scaffolds: Enhancing biocompatibility and osteoimmunomodulation. Colloids and Surfaces B: Biointerfaces 2024;234:113717. https://doi.org/10.1016/j.colsurfb.2023.113717.

[53] Matienzo LJ, Winnacker SK. Dry Processes for Surface Modification of a Biopolymer: Chitosan. Macromolecular Materials and Engineering 2002;287:871–80. https://doi.org/10.1002/mame.200290022.

[54] Maachou H, Genet MJ, Aliouche D, Dupont-Gillain CC, Rouxhet PG. XPS analysis of chitosan–hydroxyapatite biomaterials: from elements to compounds. Surface and Interface Analysis 2013;45:1088–97. https://doi.org/10.1002/sia.5229.

[55] Stevens JS, Luca ACD, Pelendritis M, Terenghi G, Downes S, Schroeder SLM. Quantitative analysis of complex amino acids and RGD peptides by X-ray photoelectron spectroscopy (XPS). Surface and Interface Analysis 2013;45:1238–46. https://doi.org/10.1002/sia.5261.

[56] Lawrie G, Keen I, Drew B, Chandler-Temple A, Rintoul L, Fredericks P, et al. Interactions between Alginate and Chitosan Biopolymers Characterized Using FTIR and XPS. Biomacromolecules 2007;8:2533–41. https://doi.org/10.1021/bm070014y.

[57] Kurmaev EZ, Shin S, Watanabe M, Eguchi R, Ishiwata Y, Takeuchi T, et al. Probing oxygen and nitrogen bonding sites in chitosan by X-ray emission. Journal of Electron Spectroscopy and Related Phenomena 2002;125:133–8. https://doi.org/10.1016/S0368-2048(02)00094-4.

[58] Wang Z, Sun C, Vegesna G, Liu H, Liu Y, Li J, et al. Glycosylated aniline polymer sensor: Amine to imine conversion on protein–carbohydrate binding. Biosensors and Bioelectronics 2013;46:183–9. https://doi.org/10.1016/j.bios.2013.02.030.

[59] Liu Y, Liu Y, Park M, Park SJ, Zhang Y, Akanda MR, et al. Green synthesis of fluorescent carbon dots from carrot juice for in vitro cellular imaging. Carbon Letters 2017;21:61–7. https://doi.org/10.5714/CL.2017.21.061.

[60] Maerten C, Garnier T, Lupattelli P, Chau NTT, Schaaf P, Jierry L, et al. Morphogen Electrochemically Triggered Self-Construction of Polymeric Films Based on Mussel-Inspired Chemistry. Langmuir 2015;31:13385–93. https://doi.org/10.1021/acs.langmuir.5b03774.

[61] Ando RA, Nascimento GMD, Landers R, Santos PS. Spectroscopic investigation of conjugated polymers derived from nitroanilines. Spectrochimica Acta Part A: Molecular and Biomolecular Spectroscopy 2008;69:319–26. https://doi.org/10.1016/j.saa.2007.03.046.

[62] Stevens JS, Schroeder SLM. X-Ray Photoelectron Spectroscopy. Encyclopedia of Physical Organic Chemistry, John Wiley & Sons, Ltd; 2017, p. 1–53. https://doi.org/10.1002/9781118468586.epoc4031.

[63] Prabhuram J, Wang X, Hui CL, Hsing IM. Synthesis and characterization of surfactant-stabilized Pt/C nanocatalysts for fuel cell applications. Journal of Physical Chemistry B 2003;107:11057–64. https://doi.org/10.1021/jp0357929.

[64] Takeuchi M, Shimizu Y, Yamagawa H, Nakamuro T, Anpo M. Preparation of the visible light responsive N3−-doped WO3 photocatalyst by a thermal decomposition of ammonium paratungstate. Applied Catalysis B: Environmental 2011;110:1–5. https://doi.org/10.1016/j.apcatb.2011.08.004.

[65] Remy MJ, Genet MJ, Lardinois PF, Notté PP, Poncelet G. Acidity of the outer surface of dealuminated mordenites as studied by X-ray photoelectron spectroscopy of chemisorbed ammonia. Surface and Interface Analysis 1994;21:643–9. https://doi.org/10.1002/sia.740210908.

[66] Simon-Walker R, Romero R, Staver JM, Zang Y, Reynolds MM, Popat KC, et al. Glycocalyx-Inspired Nitric Oxide-Releasing Surfaces Reduce Platelet Adhesion and Activation on Titanium. ACS Biomater Sci Eng 2017;3:68–77. https://doi.org/10.1021/acsbiomaterials.6b00572.

[67] Amornsudthiwat P, Mongkolnavin R, Kanokpanont S, Panpranot J, Wong CS, Damrongsakkul S. Improvement of early cell adhesion on Thai silk fibroin surface by low energy plasma. Colloids and Surfaces B: Biointerfaces 2013;111:579–86. https://doi.org/10.1016/j.colsurfb.2013.07.009.

[68] Jedlicka SS, Rickus JL, Zemlyanov DY. Surface Analysis by X-ray Photoelectron Spectroscopy of Sol−Gel Silica Modified with Covalently Bound Peptides. J Phys Chem B 2007;111:11850–7. https://doi.org/10.1021/jp0744230.

[69] Zheng X, Ke X, Yu P, Wang D, Pan S, Yang J, et al. A facile strategy to construct silk fibroin based GTR membranes with appropriate mechanical performance and enhanced osteogenic capacity. J Mater Chem B 2020;8:10407–15. https://doi.org/10.1039/D0TB01962C.

[70] Pantea D, Darmstadt H, Kaliaguine S, Roy C. Electrical conductivity of conductive carbon blacks: influence of surface chemistry and topology. Applied Surface Science 2003;217:181–93. https://doi.org/10.1016/S0169-4332(03)00550-6.

[71] Hsu LC, Fang J, Borca-Tasciuc DA, Worobo RW, Moraru CI. Effect of micro- and nanoscale topography on the adhesion of bacterial cells to solid surfaces. Applied and Environmental Microbiology 2013;79:2703–12. https://doi.org/10.1128/AEM.03436-12.

[72] Cordero M, Mitarai N, Jauffred L. Motility mediates satellite formation in confined biofilms. ISME J 2023;17:1819–27. https://doi.org/10.1038/s41396-023-01494-x.

[73] Gülez G, Dechesne A, Smets B. The Pressurized Porous Surface Model: An improved tool to study bacterial behavior under a wide range of environmentally relevant matric potentials. Journal of Microbiological Methods 2010;82:324–6. https://doi.org/10.1016/j.mimet.2010.06.009.

[74] Friedlander RS, Vlamakis H, Kim P, Khan M, Kolter R, Aizenberg J. Bacterial flagella explore microscale hummocks and hollows to increase adhesion. Proceedings of the National Academy of Sciences 2013;110:5624–9. https://doi.org/10.1073/pnas.1219662110.

[75] Jin X, Marshall JS. Mechanics of biofilms formed of bacteria with fimbriae appendages. PLOS ONE 2020;15:e0243280. https://doi.org/10.1371/journal.pone.0243280.

[76] Bjarnsholt T. Introduction to Biofilms. In: Bjarnsholt T, Jensen PØ, Moser C, Høiby N, editors. Biofilm Infections, New York, NY: Springer; 2011, p. 1–9. https://doi.org/10.1007/978-1-4419-6084-9_1.

[77] Seo H, Yu X, Tripathi A, Champion JA, Harris TAL. Enhancing Bacterial Adhesion with Hydro-Softened Chitosan Films. ACS Macro Lett 2025;14:1155–61. https://doi.org/10.1021/acsmacrolett.5c00374.

[78] Srinivasan R, Santhakumari S, Poonguzhali P, Geetha M, Dyavaiah M, Xiangmin L. Bacterial Biofilm Inhibition: A Focused Review on Recent Therapeutic Strategies for Combating the Biofilm Mediated Infections. Front Microbiol 2021;12. https://doi.org/10.3389/fmicb.2021.676458.

[79] Cardinali MA, Pierantoni DC, Comez L, Conti A, Chiesa I, Maria CD, et al. Black phosphorus/silk fibroin films hamper filamentous and invasive growth of Candida albicans. RSC Adv 2024;14:39112–21. https://doi.org/10.1039/D4RA05126B.

[80] Horsthemke M, Bachg AC, Groll K, Moyzio S, Müther B, Hemkemeyer SA, et al. Multiple roles of filopodial dynamics in particle capture and phagocytosis and phenotypes of Cdc42 and Myo10 deletion. J Biol Chem 2017;292:7258–73. https://doi.org/10.1074/jbc.M116.766923.

[81] Möller J, Lühmann T, Chabria M, Hall H, Vogel V. Macrophages lift off surface-bound bacteria using a filopodium-lamellipodium hook-and-shovel mechanism. Sci Rep 2013;3:2884. https://doi.org/10.1038/srep02884.

[82] Lam J, Herant M, Dembo M, Heinrich V. Baseline Mechanical Characterization of J774 Macrophages. Biophysical Journal 2009;96:248–54. https://doi.org/10.1529/biophysj.108.139154.

[83] Smith TD, Tse MJ, Read EL, Liu WF. Regulation of macrophage polarization and plasticity by complex activation signals. Integr Biol (Camb) 2016;8:946–55. https://doi.org/10.1039/c6ib00105j.

[84] Kasemo B. Biological surface science. Surface Science 2002;500:656–77. https://doi.org/10.1016/S0039-6028(01)01809-X.

[85] Kocourková K, Kadlečková M, Wrzecionko E, Mikulka F, Knechtová E, Černá P, et al. Silk Fibroin Surface Engineering Using Phase Separation Approaches for Enhanced Cell Adhesion and Proliferation. ACS Appl Mater Interfaces 2025;17:13702–12. https://doi.org/10.1021/acsami.5c00874.

[86] Francius G, Polyakov P, Merlin J, Abe Y, Ghigo J-M, Merlin C, et al. Bacterial Surface Appendages Strongly Impact Nanomechanical and Electrokinetic Properties of Escherichia coli Cells Subjected to Osmotic Stress. PLOS ONE 2011;6:e20066. https://doi.org/10.1371/journal.pone.0020066.

[87] Alkhalifa S, Jennings MC, Granata D, Klein M, Wuest WM, Minbiole KPC, et al. Analysis of the Destabilization of Bacterial Membranes by Quaternary Ammonium Compounds: A Combined Experimental and Computational Study. ChemBioChem 2020;21:1510–6. https://doi.org/10.1002/cbic.201900698.

[88] Cheng Y, Feng G, Moraru CI. Micro- and Nanotopography Sensitive Bacterial Attachment Mechanisms: A Review. Front Microbiol 2019;10. https://doi.org/10.3389/fmicb.2019.00191.

[89] Jiang W, Saxena A, Song B, Ward BB, Beveridge TJ, Myneni SCB. Elucidation of Functional Groups on Gram-Positive and Gram-Negative Bacterial Surfaces Using Infrared Spectroscopy. Langmuir 2004;20:11433–42. https://doi.org/10.1021/la049043+.

[90] Oh JK, Yegin Y, Yang F, Zhang M, Li J, Huang S, et al. The influence of surface chemistry on the kinetics and thermodynamics of bacterial adhesion. Sci Rep 2018;8:17247. https://doi.org/10.1038/s41598-018-35343-1.

[91] Kovari DT, Wei W, Chang P, Toro J-S, Beach RF, Chambers D, et al. Frustrated Phagocytic Spreading of J774A-1 Macrophages Ends in Myosin II-Dependent Contraction. Biophys J 2016;111:2698–710. https://doi.org/10.1016/j.bpj.2016.11.009.

[92] Walkey CD, Olsen JB, Guo H, Emili A, Chan WCW. Nanoparticle Size and Surface Chemistry Determine Serum Protein Adsorption and Macrophage Uptake. J Am Chem Soc 2012;134:2139–47. https://doi.org/10.1021/ja2084338.

[93] Rizzello L, Sorce B, Sabella S, Vecchio G, Galeone A, Brunetti V, et al. Impact of Nanoscale Topography on Genomics and Proteomics of Adherent Bacteria. ACS Nano 2011;5:1865–76. https://doi.org/10.1021/nn102692m.

[94] Kong M, Chen XG, Xing K, Park HJ. Antimicrobial properties of chitosan and mode of action: A state of the art review. International Journal of Food Microbiology 2010;144:51–63. https://doi.org/10.1016/j.ijfoodmicro.2010.09.012.

[95] Sahariah P, Másson M. Antimicrobial Chitosan and Chitosan Derivatives: A Review of the Structure–Activity Relationship. Biomacromolecules 2017;18:3846–68. https://doi.org/10.1021/acs.biomac.7b01058.

[96] Geiger B, Spatz JP, Bershadsky AD. Environmental sensing through focal adhesions. Nat Rev Mol Cell Biol 2009;10:21–33. https://doi.org/10.1038/nrm2593.

[97] Mogilner A, Rubinstein B. The Physics of Filopodial Protrusion. Biophysical Journal 2005;89:782–95. https://doi.org/10.1529/biophysj.104.056515.

[98] Chau DDL, Yung KWY, Chan WWL, An Y, Hao Y, Chan HYE, et al. Attenuation of amyloid-β generation by atypical protein kinase C-mediated phosphorylation of engulfment adaptor PTB domain containing 1 threonine 35. FASEB Journal 2019;33:12019–35. https://doi.org/10.1096/fj.201802825RR.

[99] Kuo J-C. Mechanotransduction at focal adhesions: integrating cytoskeletal mechanics in migrating cells. J Cell Mol Med 2013;17:704–12. https://doi.org/10.1111/jcmm.12054.

[100] Roach P, Farrar D, Perry CC. Surface Tailoring for Controlled Protein Adsorption: Effect of Topography at the Nanometer Scale and Chemistry. J Am Chem Soc 2006;128:3939–45. https://doi.org/10.1021/ja056278e.

[101] Wang X, Li Z, Li H, Ruan S, Gu J. Influence of nanoscale surface curvature of rutile on fibronectin adsorption by atomistic simulations. J Mater Sci 2017;52:13512–21. https://doi.org/10.1007/s10853-017-1458-y.

[102] Chen C, Zhang Y, Zhang L, Ullah I, Hang L, Liu Y, et al. Exosome-functionalized photocrosslinked GelMA/HAMA hydrogel promotes facial nerve recovery via inflammatory microenvironment regulation. Bioactive Materials 2026;60:1–19. https://doi.org/10.1016/j.bioactmat.2026.01.008.

[103] Wold S, Sjöström M, Eriksson L. PLS-regression: a basic tool of chemometrics. Chemometrics and Intelligent Laboratory Systems 2001;58:109–30. https://doi.org/10.1016/S0169-7439(01)00155-1.

[104] Wold S, Kettaneh N, Tjessem K. Hierarchical multiblock PLS and PC models for easier model interpretation and as an alternative to variable selection. Journal of Chemometrics 1996;10:463–82. https://doi.org/10.1002/(SICI)1099-128X(199609)10:5/6%253C463::AID-CEM445%253E3.0.CO;2-L.

[105] Kuligowski J, Pérez-Rubio Á, Moreno-Torres M, Soluyanova P, Pérez-Rojas J, Rienda I, et al. Cluster-Partial Least Squares (c-PLS) regression analysis: Application to miRNA and metabolomic data. Analytica Chimica Acta 2024;1286:342052. https://doi.org/10.1016/j.aca.2023.342052.

Supplementary Information for:

# Multiscale Analysis of Plasma-Modified Silk Fibroin and Chitosan Films

Jordan Nashed*[a,b], Tomasz Bartkowiak[c], Alexandru Horia Marin[a,d], Tine Curk[b], and Viviana Marcela Posada-Perez*[a,e]

[a]Ken and Mary Alice Lindquist Department of Nuclear Engineering, Pennsylvania State University, PA USA

[b]Department of Materials Science and Engineering, Johns Hopkins University, MD USA

[c]Institute of Mechanical Technology, Poznań University of Technology, plac Marii Skłodowskiej-Curie 5, 60-965 Poznań, Poland

[d]Materials Science and Engineering Department, Pennsylvania State University, University Park, PA, 16802, USA

[e]Department of Electronics and Computer Science, Pontificia Universidad Javeriana de Cali, Colombia

[f]Surface Analysis Laboratory, Institute for Nuclear Research Pitesti, Mioveni, 15400, Romania

*Corresponding authors

**1.** Single-Scale Topographic Data

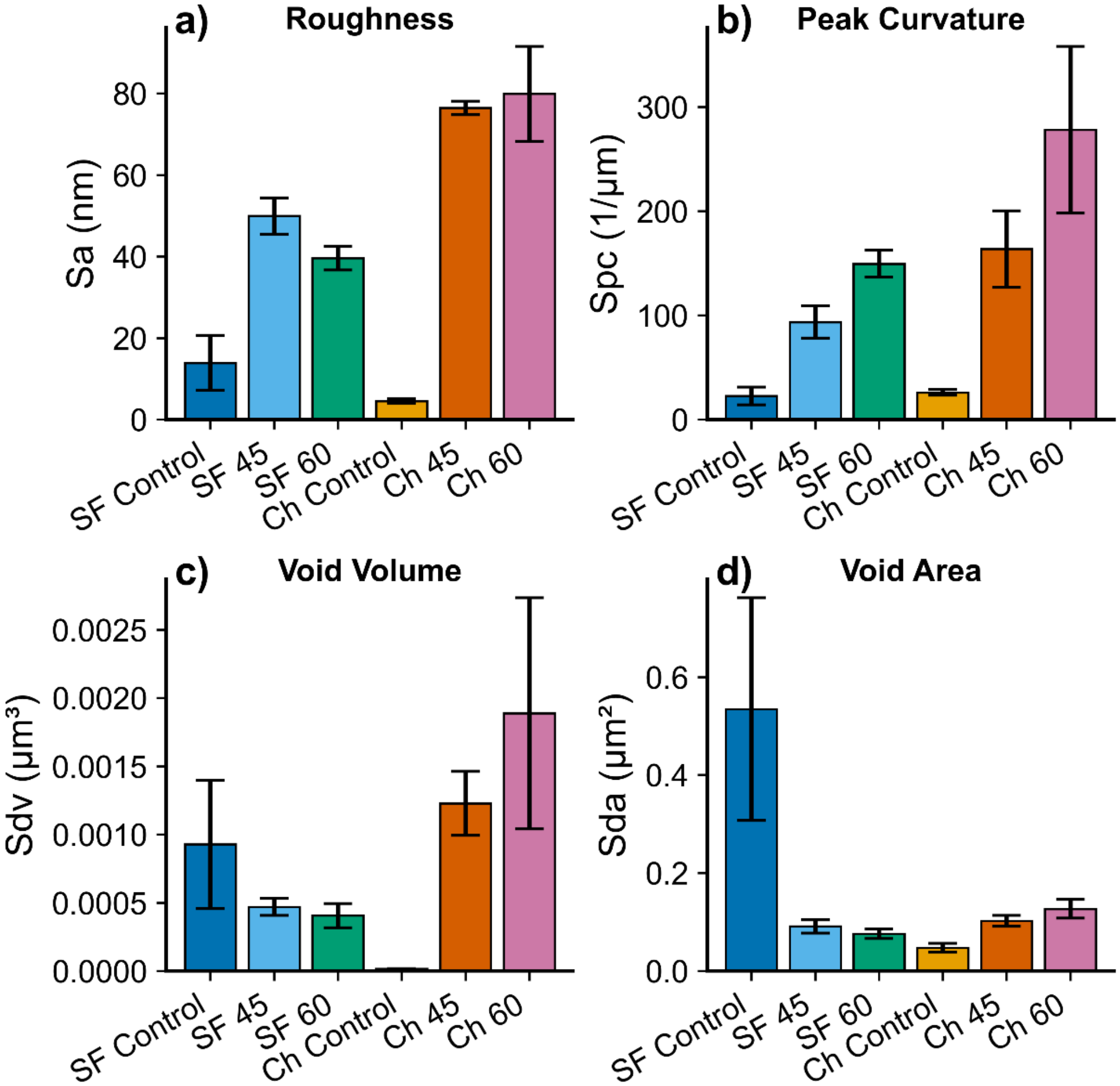

**Supplementary Figure S1. Representative conventional (single-scale) ISO topographic parameter values for treated and control samples.** Single-scale data was simply measured from unfiltered AFM images using the same parameters and family designations as in the multiscale bandpass analysis. Error bars indicate the standard error of the five samples per material-treatment group. a) Sa: arithmetic mean height, b) Spc: mean peak curvature, c) Sdv: mean dale volume, and d) Sda: mean dale area.

**2.** Bacterial Colony Size Distribution

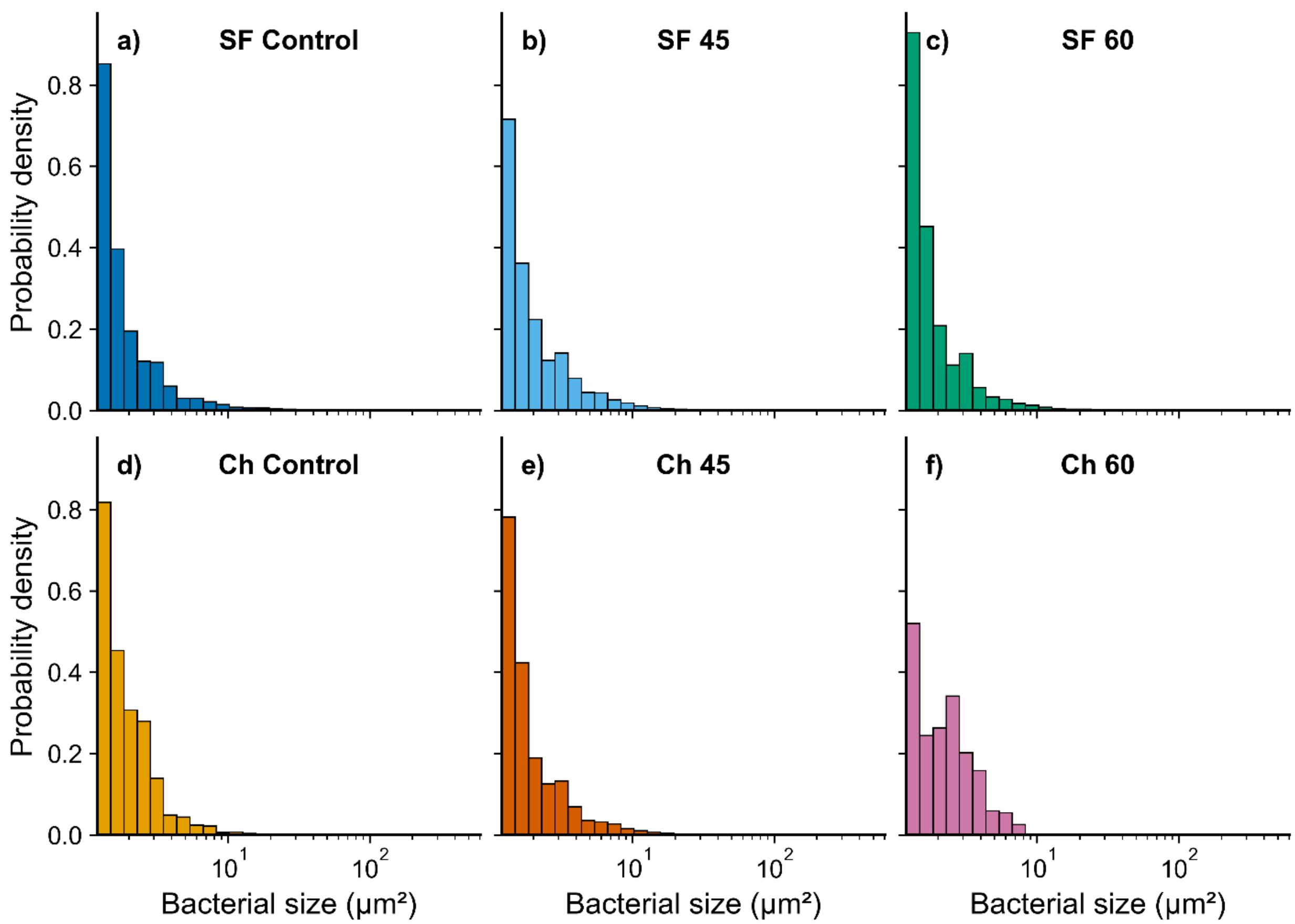

**Supplementary Figure S2: Aggregated size distribution of bacterial colonies on each material-treatment group, shown on logarithmic scale for clarity.** a) SF Control, b) SF 45, c) SF 60, d) Ch Control, e) Ch 45, and f) Ch 60.

**3.** Functional Group Abundance

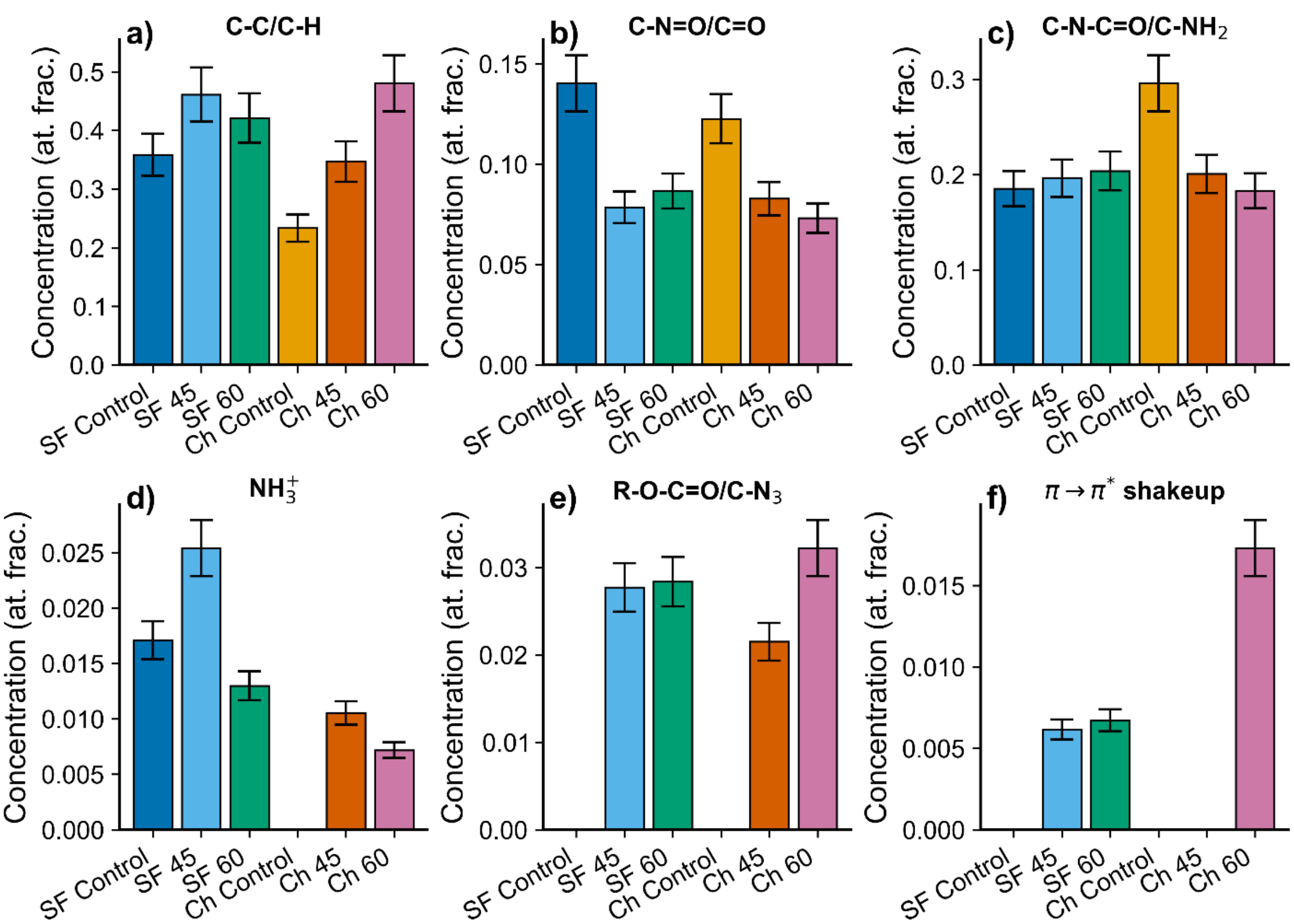

**Supplementary Figure S3:** Functional group abundances from XPS. Error bars represent 10% concentration due to estimated error in measured concentration values. All functional group abundances other than $NH_3^+$ (N1s peak) were derived from the C1s peak. a) C-C/C-H, b) C-N=O/C=O, c) C-N-C=O/C-$NH_2$, d) $NH_3^+$, e) R-O-C=O/C-$N_3$, f) pi to pi* shakeup. Low overall abundance of $NH_3^+$, R-O-C=O/C-$N_3$, and pi to pi* shakeup indicate that a causal relationship between their concentrations and the biological response is unlikely, reflected in the analysis and discussion of the heatmaps in the main text.

**4.** Discriminating Surfaces: Principal Component Analysis

To demonstrate that multiscale surface descriptors capture treatment-specific signatures that are not resolved by single-scale metrics, principal component analysis (PCA) was employed to assess the ability of the measured surface properties to differentiate between the six material–treatment groups. PCA of the material data helped elucidate the discriminating surface features defining the treatment/material groups for both single-scale and multiscale datasets (**Supplementary Figure S4**). Closeness in this space indicates a high degree of similarity in the measured surface parameters.

PCA was expectedly better able to distinguish between groups using multiscale data than single-scale data, as indicated by the increased separation between the six groups in the plots, particularly for the 45-degree irradiated samples. PC1 captured much of the variance in surface data and was the most discriminatory for both single- and multiscale datasets. Therefore, PC1 loadings were investigated to understand the essential latent structure encoded by PC1 and thus the latent structure differentiating the material surfaces (**Supplementary Figure S5**).

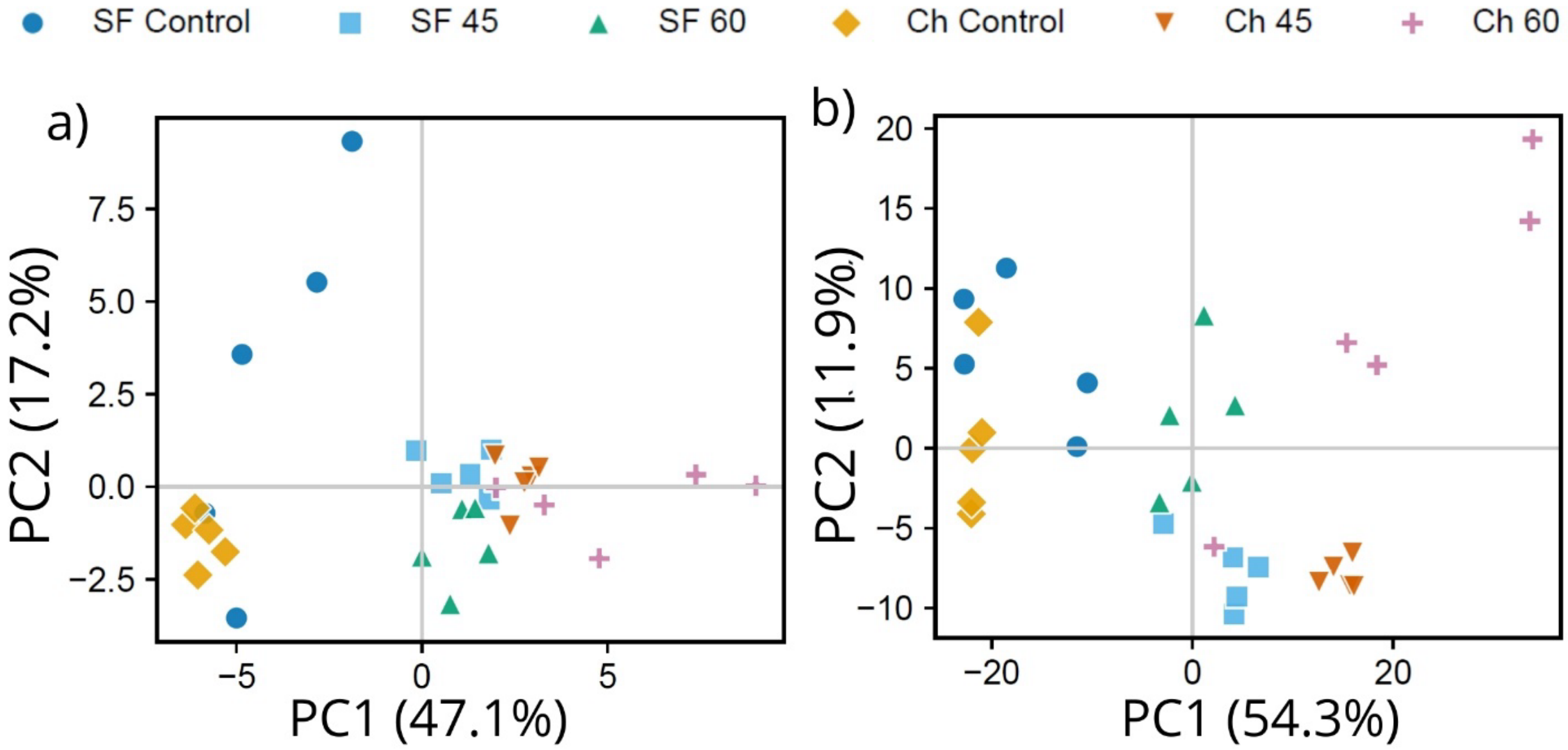

**Supplementary Figure S4. PCA score plots for (A) single-scale and (B) multiscale surface-parameter datasets with captured variance (%) for each principal component listed.** The distributions highlight how control, 45°, and 60° treatments cluster for both silk fibroin (SF) and chitosan (Ch) samples, showing the separation driven by angle-dependent topographic features.

Principal component 1 (PC1) captured nearly half the variance in both datasets. A higher value of PC1 indicates increased nanostructuring and is a proxy for DPNS treatment.

PC1 loadings are shown in **Supplementary Figure S5**. Single-scale data had positive loadings for parameters associated with nanostructuring (i.e., roughness, volume, area, and complexity), ester/azide concentration, and (correcting for sign conventions) anisotropy and void curvature. Notable negative loadings were associated with amide/carbonyl concentration. Therefore, positive PC1 scores were indicative of higher degrees of nanostructuring and a relative increase in ester/azide concentration compared to amide/carbonyl concentration. Amide/carbonyl concentration decreased with treatment, while ester/azide concentration as well as nanostructuring increased with treatment (**Supplementary Figure S3**). Therefore, PC1 was effectively a proxy for DPNS treatment. Multiscale loadings exhibited similar trends; factors associated with nanostructuring dominated the positive loadings and ester/azide concentration was opposite the loading for amide/carbonyl concentration. Loadings were largely uniformly distributed along scales for each given parameter, likely due to strong multicollinearity and the large number of variables.

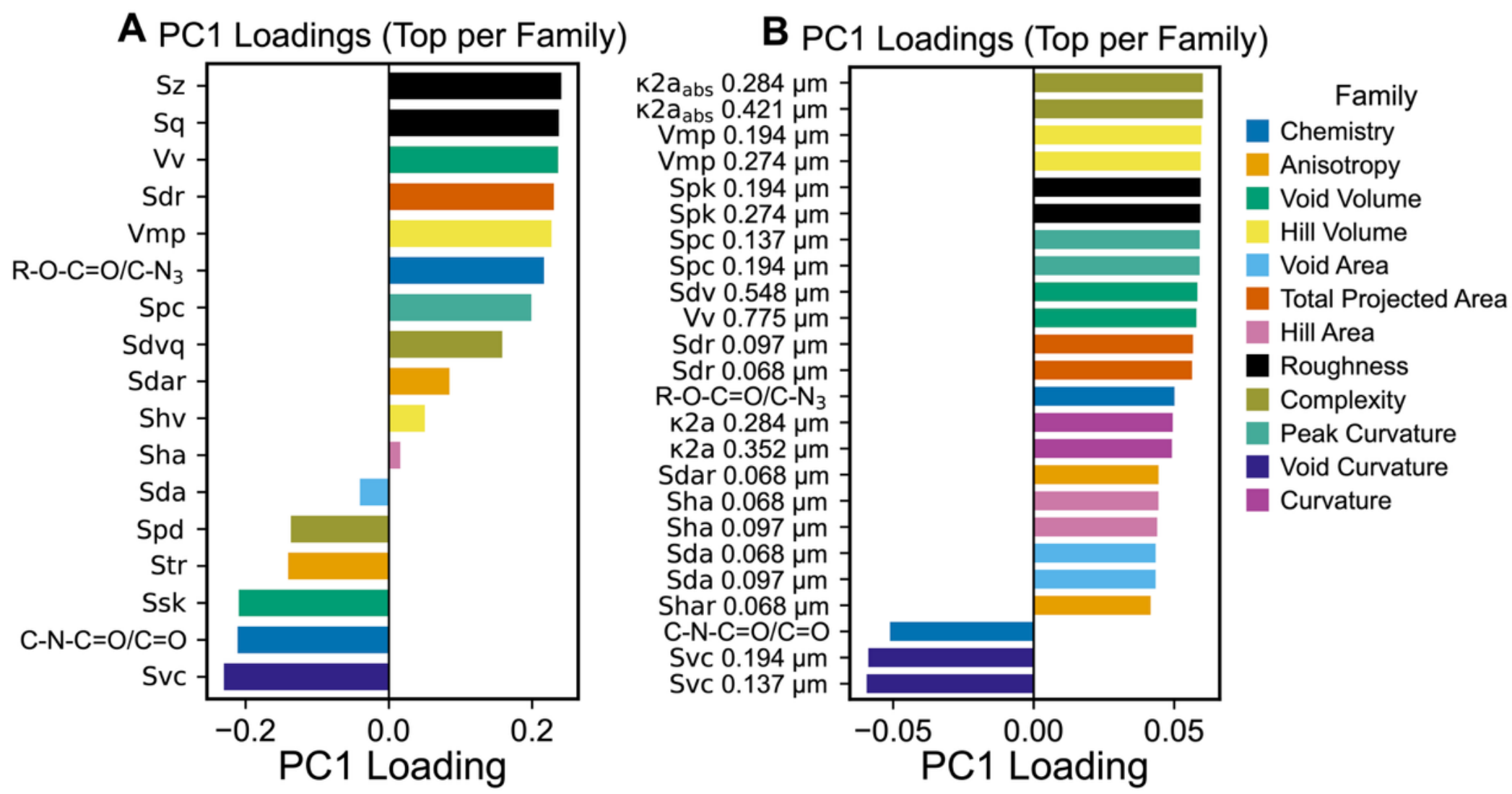

**Supplementary Figure S5. PC1 loadings for the top two features within each functional family.** The left panel (A) shows the highest-loading parameters from the multiscale dataset (502 total parameters), while the right panel (B) shows the corresponding top features from the conventional (single-scale) dataset (35 total parameters). Colors denote the functional family associated with each parameter.

Importantly, PC1 emerged as a latent multiscale surface signature rather than a simple roughness index, integrating curvature, void structure, anisotropy, and complexity across scales. The relatively uniform distribution of loadings reflects strong multicollinearity among scale-resolved descriptors and suggests hierarchical surface restructuring rather than dominance of a single scale. In this sense, multiscale analysis does not introduce redundant information but reveals coordinated reorganization of surface features across length scales. Although chemical parameters contributed less to the total variance, their alignment with PC1 indicates that near-surface chemical changes follow the same multiscale trend, as further examined by XPS in the following section.

## 5. Multiscale Curvature Tensor Heat Maps

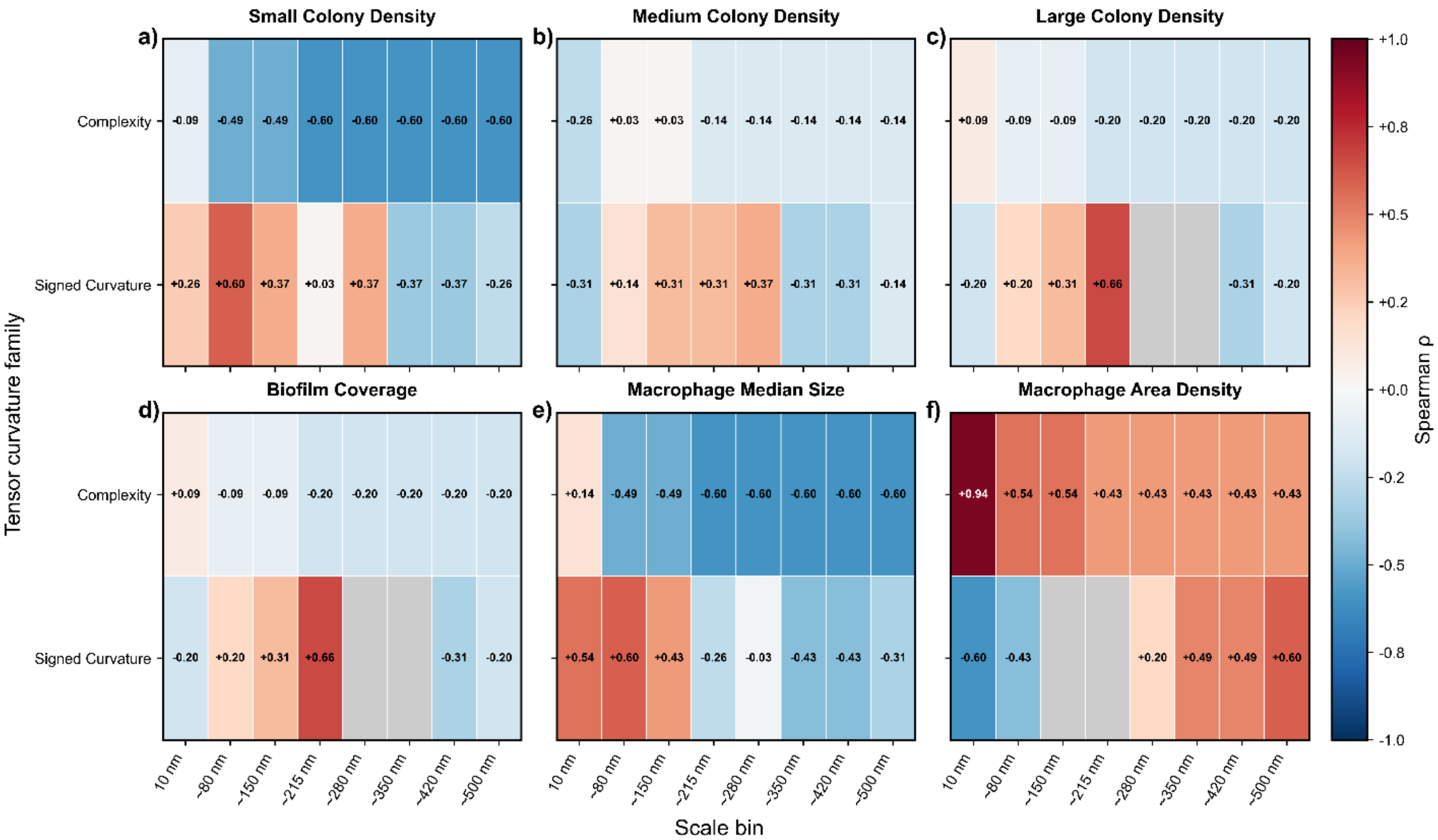

**Supplementary Figure S6: Multiscale curvature tensor heat maps.** a) Small colony density, b) medium colony density, c) large colony density, d) biofilm coverage, e) macrophage median size, and f) macrophage area density. Small colony density, macrophage median size, and macrophage area density showed the strongest correlations, mirroring bandpass results. Differences in scale of correlation between bandpass and tensor data reflect the fundamental differences in scale definitions (central wavelength vs mesh spacing) and topographic parameters.

**6.** Multiscale Data Relative Scores

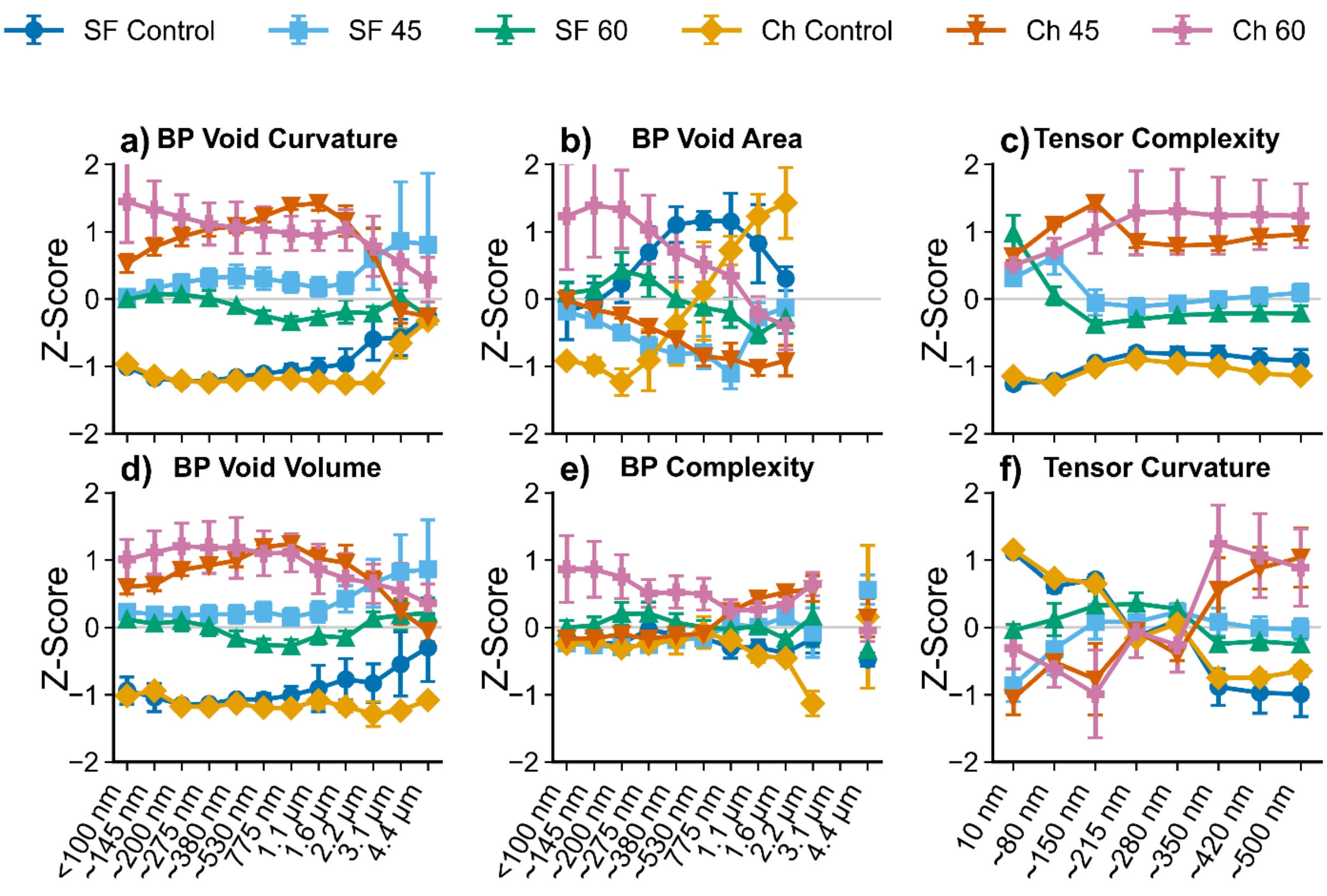

**Supplementary Figure S7: Multiscale relative scores (Z-scores) showing relative values of topographic families.** Error bars indicate standard error of the Z-scores for the five samples per material-treatment group. a) Bandpass void curvature, b) bandpass void area, c) multiscale tensor complexity, d) bandpass void volume, e) bandpass topographic complexity, and f) Multiscale tensor signed curvature. Treated Ch shows consistently higher scores in topographic parameters, suggesting more pronounced topographic changes relative to treated SF. Signed curvature (f) shows an inversion in relative scores with decreasing scale, indicating increasing valley character for treated materials with decreasing scale, in line with multiscale bandpass data and multiscale curvature tensor heatmaps.

# 7. Appendix A

## 7.1. Table A.1: Multiscale Curvature Tensor Parameters and Family Assignments

| **Complexity**[1] | |
|---|---|
| Symbol: | Description: |
| Kq | RMS Gaussian curvature |
| Hq | RMS mean curvature |
| $Kq_{abs}$ | RMS absolute Gaussian curvature |
| $Ka_{abs}$ | Average absolute Gaussian curvature |
| $Ha_{abs}$ | Mean absolute curvature |
| $Hq_{abs}$ | RMS mean absolute curvature |
| κ1q | RMS maximum principal curvature |
| κ2q | RMS minimum principal curvature |
| $\kappa 1q_{abs}$ | RMS maximum absolute principal curvature |
| $\kappa 2q_{abs}$ | RMS minimum absolute principal curvature |
| $\kappa 1a_{abs}$ | Mean maximum absolute principal curvature |
| $\kappa 2a_{abs}$ | Mean minimum absolute principal curvature |
| **Signed Curvature** | |
| Symbol: | Description: |
| Ha | Mean curvature |
| κ1a | Mean maximum principal curvature |
| κ2a | Mean minimum principal curvature |

[1] Complexity here broadly refers to the change of material properties with scale as well as surface irregularity.

## 7.2. Table A.2: Bandpass Parameters and Family Assignments

| **Anisotropy** | |
| --- | --- |
| Symbol: | Description: |
| Shar | Mean hill aspect ratio |
| Sdar | Mean dale[2] aspect ratio |
| Str[3] | Mean texture aspect ratio |
| **Void Volume** | |
| Symbol: | Description: |
| Vv | Void volume |
| Sdv | Mean dale volume |
| Ssk[4] | Skewness of height distribution |
| **Hill Volume** | |
| Symbol: | Description: |
| Shv | Mean hill volume |
| Vmp | Peak material volume |
| **Void Area** | |
| Symbol: | Description: |
| Sda | Mean dale area |
| **Total Projected Area** | |
| Symbol: | Description: |
| Sdr | Developed interfacial area ratio (ratio of actual surface to projected area) |

[2] Dale is the terminology used in ISO 25178; void is used in place of dale in the main text for clarity.
[3] An Str value of 1 indicates a purely isotropic surface; therefore, 1 – Str was used for anisotropy.
[4] Negative Ssk indicates higher void character; therefore, -Ssk was used for void volume.

| Hill Area | |
|---|---|
| Symbol: | Description: |
| Sha | Mean hill area |
| **Roughness** | |
| Symbol: | Description: |
| Sp | Maximum peak height |
| Sv | Maximum void depth |
| Sz | Maximum height |
| Sa | Arithmetic mean height (areal mean roughness) |
| Sq | RMS height (areal RMS roughness) |
| Spk | Reduced peak height (mean height of peaks above core surface) |
| Svk | Reduced valley depth (mean depth of valleys below core surface) |
| **Complexity** | |
| Symbol: | Description: |
| Shaq | Standard deviation in mean hill area |
| Sdaq | Standard deviation in mean dale area |
| Sharq | Standard deviation in mean hill aspect ratio |
| Sdarq | Standard deviation in mean dale aspect ratio |
| Sku | Height distribution kurtosis[5] |
| Shvq | Standard deviation in mean hill volume |
| Sdvq | Standard deviation in mean dale volume |
| Spd | Mean peak area density |
| Svd | Mean void area density |

[5] Sku above 3 indicates more 'extremes' (longer tails) than a normal distribution; therefore, Sku – 3 was used to measure 'excess' height complexity.

| Peak Curvature | |
|---|---|
| Symbol: | Description: |
| Spc | Mean peak curvature |
| **Void Curvature** | |
| Symbol: | Description: |
| Svc | Mean void curvature |